\documentclass{emulateapj}
\usepackage{txfonts}
\usepackage{graphicx}

\slugcomment{to appear in special issue ApJSS., 2016 }

\shorttitle{X-ray flux and spectral parameter uncertainties from {\it Chandra} ACIS-I spectra}

\shortauthors{Albacete-Colombo et al.}

\begin{document}


\title{The statistical uncertainties on X-ray flux and spectral parameters from {\it Chandra} ACIS-I observations of faint sources:  
Application to the Cygnus OB2 Association}


\author{J. F. Albacete-Colombo\altaffilmark{1, 2}}
\affil{Universidad de Rio Negro, Sede Atl\'antica - CONICET, Viedma CP8500, Argentina.}
\email{e-mail: albacete.facundo@conicet.gov.ar}

\author{E. Flaccomio\altaffilmark{2}}
\affil{Osservatorio Astronomico di Palermo, Palermo, 90134, Italy.}

\author{J. J. Drake\altaffilmark{3}}
\affil{Smithsonian Astrophysical Observatory, Cambridge, USA.}

\author{N. J. Wright\altaffilmark{4}}
\affil{Astrophysics Group, Keele University, Keele, Staffordshire ST5 5BG, UK}

\author{M. Guarcello\altaffilmark{2}}
\affil{Osservatorio Astronomico di Palermo, Palermo, 90134, Italy.}

\and

\author{V. Kashyap\altaffilmark{3}}
\affil{Smithsonian Astrophysical Observatory, Cambridge, USA.}

\altaffiltext{1}{External post-doctoral fellow in the visiting programmer of CONICET.}

\begin{abstract}
We investigate the uncertainties of fitted X-ray model parameters and
fluxes for relatively faint {\it Chandra} ACIS-I source
spectra. Monte-Carlo (MC) simulations are employed to construct a
large set of 150,000 fake X-ray spectra in the low photon count
statistics regime (from 10 to 350 net counts) using the XSPEC spectral
model fitting package. The simulations employed both absorbed thermal
(APEC) and non-thermal (power-law) models, in concert with the {\it
Chandra} ACIS-I instrument response and interstellar absorption.
Simulated X-ray spectra were fit assuming a wide set of different
input parameters and C-statistic minimization criteria to avoid
numerical artifacts in the accepted solutions.  Results provide an
error estimate for each parameter (absorption, $N_{\rm H}$, plasma
temperature, $kT$, or power-law slope, $\Gamma$, and flux), and
for different background contamination levels.  The distributions of
these errors are studied as a function of the 1$\sigma$ quantiles and
we show how these correlate with different model parameters, net
counts in the spectra and relative background level.  
Maps of uncertainty in terms of the 1$\sigma$
quantiles for parameters and flux are computed as a function of
spectrum net counts.  We find very good agreement between our estimated X-ray
spectral parameter and flux uncertainties and those recovered from spectral fitting 
for a subset of the X-ray sources detected in the Chandra Cygnus OB2 Legacy Survey 
diagnosed to be Association members and that
have between 20 and 350 net counts.  Our method can provide 
uncertainties for spectral parameters whenever formal X-ray spectral fits
cannot be well-constrained, or are unavailable, and predictions 
useful for computing {\it Chandra} ACIS-I exposure times for observation planning.   
\end{abstract}


\section{Introduction}

The current generation of flagship soft X-ray telescopes, i.e. {\em Chandra} and
{\em XMM-Newton}, both now 15 years old, ushered in a new era of X-ray astronomy
thanks to their greatly improved sensitivity and spatial resolution with
respect to their predecessors. Large numbers of previously unknown
Galactic and extra-Galactic X-ray sources were discovered and their
emission studied in detail.  Most of the X-ray sources, however, have been 
detected with just a few photons, often leaving their astrophysical
nature loosely constrained by the X-ray data alone. Quite frequently, there 
are too few photon counts to distinguish between plausible emission models.
Even when the appropriate type of emission model to apply is known with reasonable 
certainty, the relevant spectral parameters, among the most important of which is 
the intrinsic absorption-corrected source flux, are often too weakly constrained to
allow useful physical inference. The degree to which X-ray emission
models can be constrained observationally depends on the instrumental
characteristics (the effective area {\em and} spectral response), on the
source flux, on the shape of the source spectrum, on the foreground and intrinsic 
extinction, and on the instrumental and astrophysical background
contribution to the collected spectrum.

Although standard spectral fitting procedures can provide estimates
of uncertainties on model parameters, it would also be useful to have a
solid a-priori understanding of uncertainties even before the
observation is performed, such as when planning new observations. Such
estimates are crucial to both determine if the astrophysical problem at
hand is tractable, and to choose the best instrument, setup, and
exposure time combination. More specifically, given an estimate of the
broad band photon flux, common questions are: $i)$ will the observed
spectra constrain the emission mechanism, i.e. will some of the possible
emission mechanisms be rejected on the basis of the spectral shape? $ii)$
which parameters can be constrained simultaneously and within a given
precision, through spectral fitting with a specific emission model?
$iii$) to what degree are the estimates of the spectral parameters
correlated (e.g.\ N$_{\rm H}$-kT or N$_{\rm H}$-$\Gamma$ for 
thermal or non-thermal emission subject to interstellar absorption)? 

Unfortunately, no appropriate tool to address these questions is available, to our
knowledge, and resorting to Monte-Carlo (MC) simulations for each specific case 
may be not be practical and can be overly time consuming.  As a
result, the preparation of new observing campaigns, and the proposal
selection process, are too often based on previous more-or-less relevant
experiences or qualitative assessment, rather than on firm quantitative estimates.
Some of these questions were discussed by \citet{Maggio1995} for the
case of the ROSAT-PSPC detector, using MC simulations 
to investigate the dependence of fit-quality with counting statistics, and, in particular,
the ability to discern isothermal from multi-temperature plasma
emission (e.g.\ one vs.\ two thermal components). With the increased
effective area and reduced background of modern-day X-ray telescopes,
the occurrence of sources detected with a handful to a
few tens of photons has increased dramatically with respect to the
times of ROSAT. Exploiting the information contained in low-count
detections, and understanding the limitations, has thus become more
pressing then ever.

In this contribution, we present the results of extensive MC
simulations intended to answer the above questions for the rather common
case of faint X-ray sources with thermal and/or power-law emission
spectra. One such case is the observation with {\em Chandra} ACIS-I of the
young OB association, Cygnus\,OB2, inevitably also containing a
population of background sources, mostly extragalactic active galactic nuclei (AGN) and a
population of foreground sources, mostly normal stars. Young stars are
expected to have rather energetic thermal emission spectra
(kT$\sim$2-4\,keV) and, in the case of Cyg\,OB2, to be considerably
absorbed by both their parent cloud and interstellar material.
Foreground stars are expected to show softer thermal spectra subject to
less absorption. Extragalactic sources will be characterized by
mostly non-thermal power-law spectra, with, on average, harder spectra
than young stars. Although it would be desirable to distinguish among
these three cases from the X-ray spectra alone, in practice, for
low-counts sources, this is often not possible.

In the following, we use MC simulations to determine the precision with
which X-ray source fluxes and model parameters can be determined from
{\em Chandra} ACIS-I observations, as a function of the model
parameters, absorption, number of photon events, and background level (the 
latter being mostly a function of exposure time and source position in the
focal plane). The paper is structured as follows: in \S\,2, we detail our
MC procedure for thermal (T) and non-thermal (NT) emission models and
our exploration of the parameter/count-statistic/background space.
We also discuss the effects on uncertainties of rather technical aspects
of spectral fitting, such as the way spectra are rebinned before
fitting. In \S\,3, we make use of our MC simulations to estimate, in the
low-background regime, the uncertainties on unabsorbed X-ray fluxes and
model parameters for each point on our three-dimensional simulation grid
(with axes: source counts, source extinction, parametrized by the
hydrogen column density N$_{\rm H}$, and a single spectral parameter for
the source, kT for thermal models or $\Gamma$ for non-thermal ones). We
also discuss how these results are interpolated to estimate
uncertainties at any point within the grid. In \S\,4, we extend
the results presented in \S\,3 to sources with non-negligible
background. Section\,5 compares our {\it a priori} estimates of uncertainties
with the results of X-ray spectral analysis of low-count stellar
sources detected in the Cyg\,OB2 star forming region as part of
the Cygnus OB2 Chandra Legacy Survey (Drake et al., this issue), 
presented as an X-ray catalog of sources \citep{Wright2014} and 
\citep{Wright2015} and an accompanying 
catalog of optical and infrared counterparts \citep{Guarcello2015}
Finally, in \S\,6 we briefly discuss the impact of our work on future science and, in
particular, for planning new observations, i.e. to compute exposures times
using the ACIS-I Chandra camera.

In the Appendix, as an example, we show a set of bi-dimensional maps 
that illustrate the application of our results to ACIS-I source spectra.\\

\section{X-ray spectral simulations}
\label{sect:xsims}

We decided to focus on two of the most commonly used X-ray emission
models: thermal emission from an optically-thin plasma and power-law
spectra, as described by the {\sc apec} \citep{Smith2001} and {\sc powerlaw} 
models within
the XSPEC\footnote{http://heasarc.gsfc.nasa.gov/docs/xanadu/xspec/index.html} 
\citep{Arnaud1996} parameter estimation code. The spectral shape of both models is fully
described by a single parameter, the plasma temperature, kT, for {\sc
apec} and the power-law index, $\Gamma$, for {\sc powerlaw}. A normalization,
that may be expressed in terms of energy flux, photon flux, or count-rate
(all in a given band) completes the description of the model.  The  {\sc apec} model 
assumes the solar mixture of heavy elements compiled by \citet{Anders1989}.  The
intrinsic source emission was absorbed using the photoelectric
absorption model {\sc tbabs}, characterized by the equivalent hydrogen
column depth N$_{\rm H}$.\\

The observed ACIS-I spectrum of a point source may be fully specified by
the incident spectrum (assumed specified by kT or $\Gamma$ for
thermal or non-thermal emission, respectively, as noted above), $N_{\rm H}$, a normalization 
factor, the instrumental response, i.e. effective area and spectral response, and by
the background contribution to the collected spectrum, of both
instrumental and astrophysical origin. In the following we will make the
simplifying assumption that, for any given incident source spectrum, the
quality of the observed spectrum, and thus the uncertainties we can
associate with model parameters after spectral fitting,  will only be 
influenced by the number of source and background photons. In other
words, we will assume that the spectral response function and the {\em
shape} of the effective area vs. energy function are the same for all
sources.   While this is not strictly true---the ACIS-I effective area depends 
weakly on the off-axis angle for example---any departures from this assumption 
are expected to have only a second order influence on the precision of derived 
spectral parameters. 

In this section, we will further assume that the background is small, not null,
typical of an almost on-axis source for which the Point Spread Function
(PSF), and thus the photon extraction area, is small. See details in section 2.1.
This restriction will be lifted in section 4. The
RMF (Response Matrix File) and ARF (Ancillary Response File) functions 
were also chosen to be appropriate for the same source. 
We will therefore only vary three parameters in the spectral simulations: $N_{\rm H}$,
kT or $\Gamma$, and the number of source counts detected in the 0.5-8.0 keV
band, net\_cnts. In a real observation, this latter quantity is the product of the
source photon flux, the exposure time, and the effective area.

Our simulation domain is defined by grids ($N_{\rm H}$, kT/$\Gamma$) that
encompass the parameter values of most astrophysical sources, as derived from
{\em Chandra} and {\em XMM-Newton} observations. In particular, for
thermal models our grid is defined by six values of N$_{\rm H}$ (0.1,
0.33, 1.0, 3.3, 10, 33.3 $\times$10$^{22}$ cm$^{-2}$) and eight values
of kT (0.5, 0.8, 1.2, 2.0, 3.0, 4.0, 6.0, 10.0\,keV) for a total of 48
combinations. For non-thermal models, we adopt the same $N_{\rm H}$ values and
five values for the power-law index $\Gamma$ (1.0, 2.0, 3.0, 4.0, 5.0),
for a total of 30 combinations.

For each point on these parameter grids, we run simulations adjusting
the model normalization so as to have, on average, spectra with a given
number of counts, $net\_cnts$, in the 0.5-8.0\,keV band. Nineteen values are 
adopted for $net\_cnts$: 10, 15, 20, 25, 30, 35, 40, 50, 60, 70, 80, 90, 100, 120, 140, 160, 200, 240, 
and 350. The number of points in our three-dimensional
grids thus amounts to 624 and 390 for thermal and non-thermal models, respectively. 
The number of simulations performed for each grid point is chosen so as to reduce 
random fluctuations on the distribution of the output parameters to an acceptable level 
(see below). A total of about 150,000 MC simulations were performed to probe the 
parameter estimation uncertainties themselves.
Another set of about 500,000 MC simulations were performed to understand the effect 
of the background contribution to the uncertainty estimation (see section 4).

All the MC simulations were performed with XSPEC v.12.8 \citep{Arnaud1996},
using our own {\it tcl} script, CIAO (v4.5) commands, and IDL\footnote{Interactive 
Data Language, \copyright Excelis Inc.} routines.
Simulated spectra, as would be observed by ACIS-I, were created starting
from our input spectral model, ARF, RMF, and background files (see
above), using the {\sc fakeit} XSPEC command, including statistical
Poisson fluctuations. Spectra were then rebinned as we would do for
real source spectra before fitting (see Sect.~\ref{sect:binning} below), and spectral fitting was
performed following the same procedure we follow for real data (ref.
Flaccomio et al. this issue). In particular, we adopt the Cash statistic
\citep[][C-stat]{Cash1979}, which is more suitable than $\chi^2$-based methods in the 
regime of small numbers of photon events, and choose as our final best-fit model
the one with the minimum C-stat value. We repeat each fit starting from several initial 
values of the model parameters $N_{\rm H}$ and kT/$\Gamma$ to reduce the chances 
of the model finding a local minimum in C-stat space.

\subsection{The role of binning in spectral fitting \label{sect:binning}}

An interesting point to address before we proceed is how the binning of
X-ray spectra along the energy axis affects the outcome of the fitting
process, and, in particular, the uncertainties on the model parameters.
A good measure of uncertainty can be obtained from the
relative error of a given parameter, which can be computed
as the ratio between the fit result and the known input value
defined in the simulated spectra.
If bin-sizes and/or  method have an effect, we should determine,
and use, an optimal strategy.

\begin{figure}[ht!]
\centering
\includegraphics[width=8.5cm,angle=0]{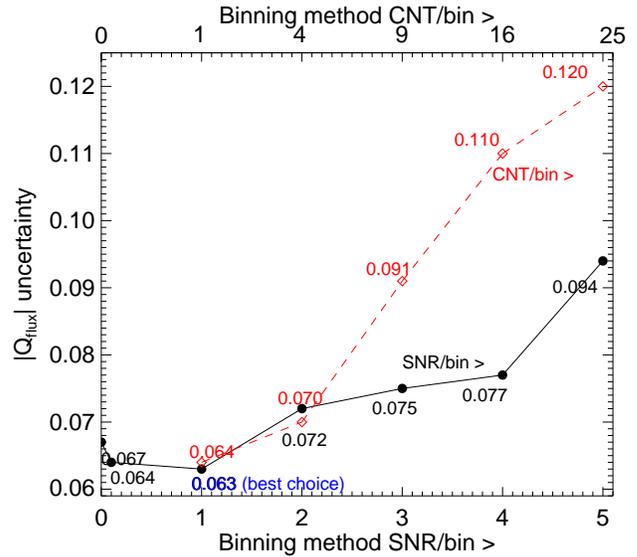}
\caption{The relative 1 $\sigma$ quantile flux errors, $|Q_{\rm flux}|$ (see text for the formal definition),  
as a function of either the minimum number of counts per bin (red points, upper X-axis) or the 
minimum SNR per bin (black points, lower X-axis) in X-ray spectral fitting simulations. 
The errors were computed from a set of 1000 simulations for each binning strategy. 
The minimum relative error occurs whenever simulated X-ray spectra are fitted with a SNR/bin$ > 1$.
The same result is derived from the analysis of relative errors of other spectral model parameters.}
\label{binning}
\end{figure}

We tested two spectral binning approaches: i) the common {\em minimum counts} method, in
which photons are grouped so that each bin includes at least the
specified number of detected photons, as implemented by the {\sc grppha} command; 
and (ii) the {\em minimum SNR} method, in which photons are grouped so as 
to reach a minimum SNR per bin, where the SNR estimation takes the
background photons into account. 
In particular, the reference background was taken to be that of a point source 
located 3.16\arcmin off-axis: the background spectrum is defined by $\sim100$ photons in the
0.5-8.0\,keV band, which must be scaled down to 0.9 photons in the source extraction 
area containing 90\% of the PSF power.
This latter method was used as implemented in the ACIS Extract/TARA IDL package 
\citep{Broos2010}; it has the added advantage that the user can specify the low and 
high-energy boundaries of the first and last bins. This is useful when fitting spectra in a
fixed energy band (in our case 0.5-8.0\,keV), because, by choosing the
proper boundaries, one can efficiently exclude photons with energies
outside that band, as opposed to having to exclude bins overlapping with
the limit boundaries and that likely contain photons with energies in
the acceptable range. This is particularly relevant for low-count
sources.

\begin{figure*}[!ht]
\centering
\includegraphics[width=18.5cm,angle=0]{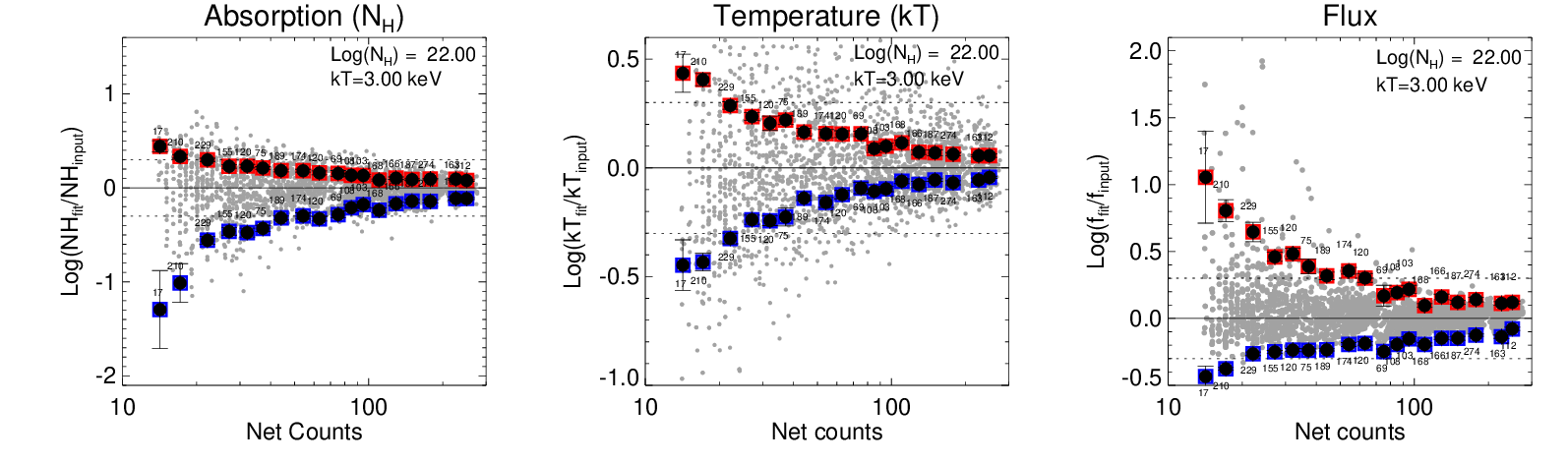}
\includegraphics[width=18.5cm,angle=0]{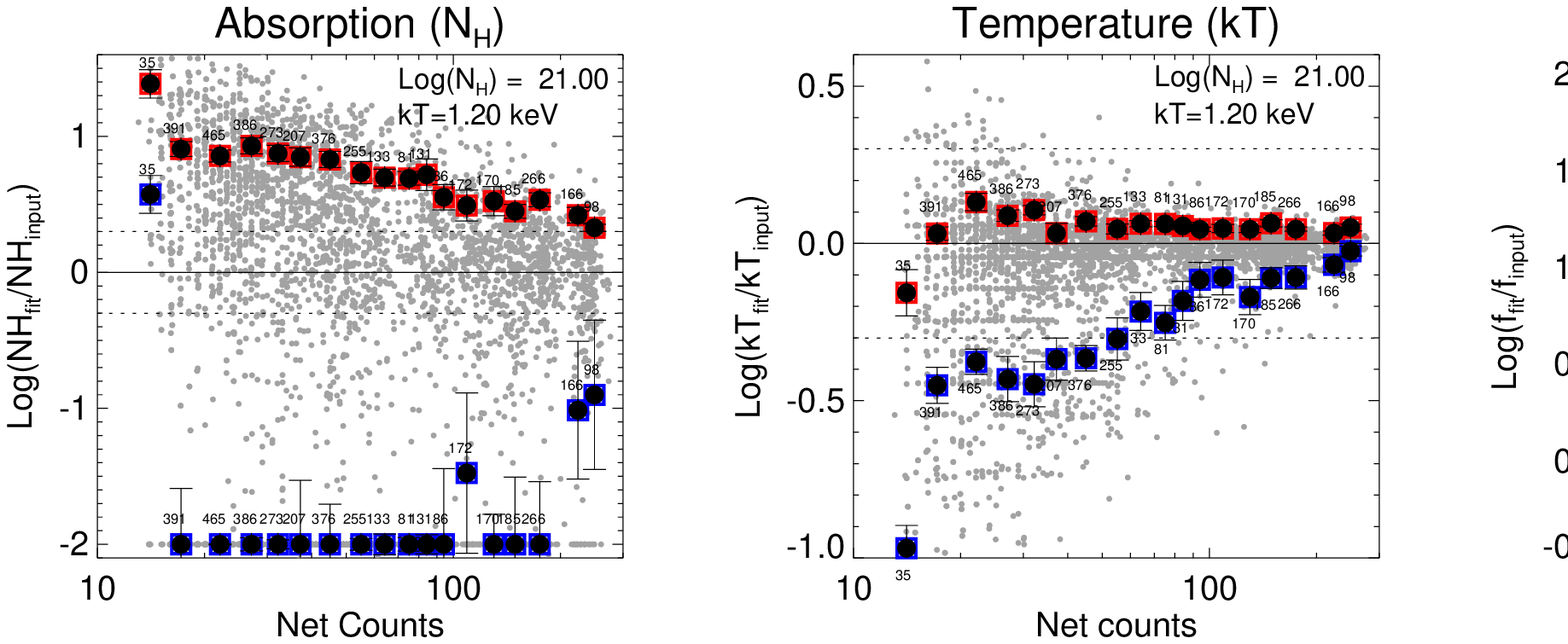}
\caption{The relative errors in N$_{\rm H}$, kT and flux as a function of the input net counts 
imposed on the spectral simulations.  The upper panels correspond to a spectral model grid 
point log($N_{\rm H}$)=22.0 and kT=3.0~keV, and the lower panels to a model with 
log($N_{\rm H}$)=21.0 and kT=1.2 keV.  The red and blue boxes refer to the 
Q-uncertainty computed for the relative error distributions (grey dots) for each incremental range 
in photon counts in which we simulate the spectra.  The numbers next to these points indicate the 
number of simulations included in each respective bin. We point out that relative errors for N$_{\rm H}$ 
in the lower central panels artificially bottom out at an $N_{\rm H}$ value a factor of 100 lower than 
the input value for cases of less than about 100 net source counts.  This numerical conditioning is 
imposed in order to avoid unconstrained lower limit values of  N$_{\rm H}$ for low column density 
cases (see also Section~\ref{s:analuncert}).} 
\label{quantiles}
\end{figure*}

We conducted our numerical experiments on binning using a specific spectral
model from our grid, an absorbed optically-thin thermal plasma spectrum with N$_{\rm
H}$=10$^{22}$\,cm$^{-2}$, kT= 2.0\,keV, and normalization set to result,
on average, in spectra with 200 net source counts. We do not expect our
conclusions to depend critically on this choice. A set of 1000
simulations was run for each binning method, i.e. for SNR$>$1, 2, 3, 4,
and 5, and for nphot$_{\rm min}$  $>$ 1, 4, 9, 16, and
25\footnote{Barring other differences in the binning algorithm,
critically the choice of limits for the first and last bins, the two
sets of values imply the same binning for the case of zero background, 
since $SNR=nphot^{1/2}$.} For each set of simulations we derive the
relative uncertainties to be expected from the spectral fit of a single
spectrum for an unabsorbed flux (0.5-8.0\,keV), N$_{\rm H}$,  and kT. 

We compute the relative error of a given Y spectral 
quantity (N$_{\rm H}$, kT / $\Gamma$, or flux) as the ratio between the parameter best-fit 
value and the model input value. Then the 1 sigma Quantile error (Q$_Y$) is an indicator 
of the uncertainty by taking the $\pm1\sigma$, i.e. 16\% (Q$^-_Y$) and 84\% (Q$^+_Y$) 
quantiles computed from the cumulative distribution of the relative errors of Y. 
Hereafter, we call such a quantity a "Q uncertainty". Its respective errors were calculated 
in order to ascertain the significance of any effect of binning parameter/method on the 
uncertainties by using the Maritz-Jarrett method \citep{Hong2004}.

Figure\,\ref{binning} shows the Q uncertainty on F$_{\rm x}$, which
were computed from a set of 1000 simulations for each binning strategy. The Q uncertainties
here are represented by the mean of the absolute value of the flux error quantiles,
$|Q_{\rm flux}| = (|Q^+_{\rm flux}| + |Q^-_{\rm flux}|)/2$.
The figure demonstrates that  Q uncertainties on best-fit parameters do actually 
depend on binning. The {\em minimum counts} method
 (CNT/bin) produces the largest uncertainties, particularly for bins with
counts/bin$>$4, while the {\em minimum SNR} method (SNR/bin) for SNR/bin$>$1 
is the best choice and less affected by statistical bias in the fitting procedure.
For large bins the superiority of the SNR method is likely accentuated by the fact that it
does not waste photons, while a growing number of photons in the bin
overlapping with the energy limits, are discarded in the former case.
Moreover, in the case of the {\em minimum counts} method, 
we find that the median values of the best-fit parameters diverge
from the synthetic spectrum input values as the binning increases, 
signalling that potentially biased results might result from such a procedure.

In summary, these numerical results indicate that binning spectra so that SNR/bin$>$1 produces
unbiased best-fit values with minimal uncertainties. This is thus the binning 
strategy we adopt in the work that follows.\\

\section{Analysis of uncertainties}
\label{s:analuncert}

\begin{figure*}[!ht]
\centering
\includegraphics[width=7.4cm,angle=0]{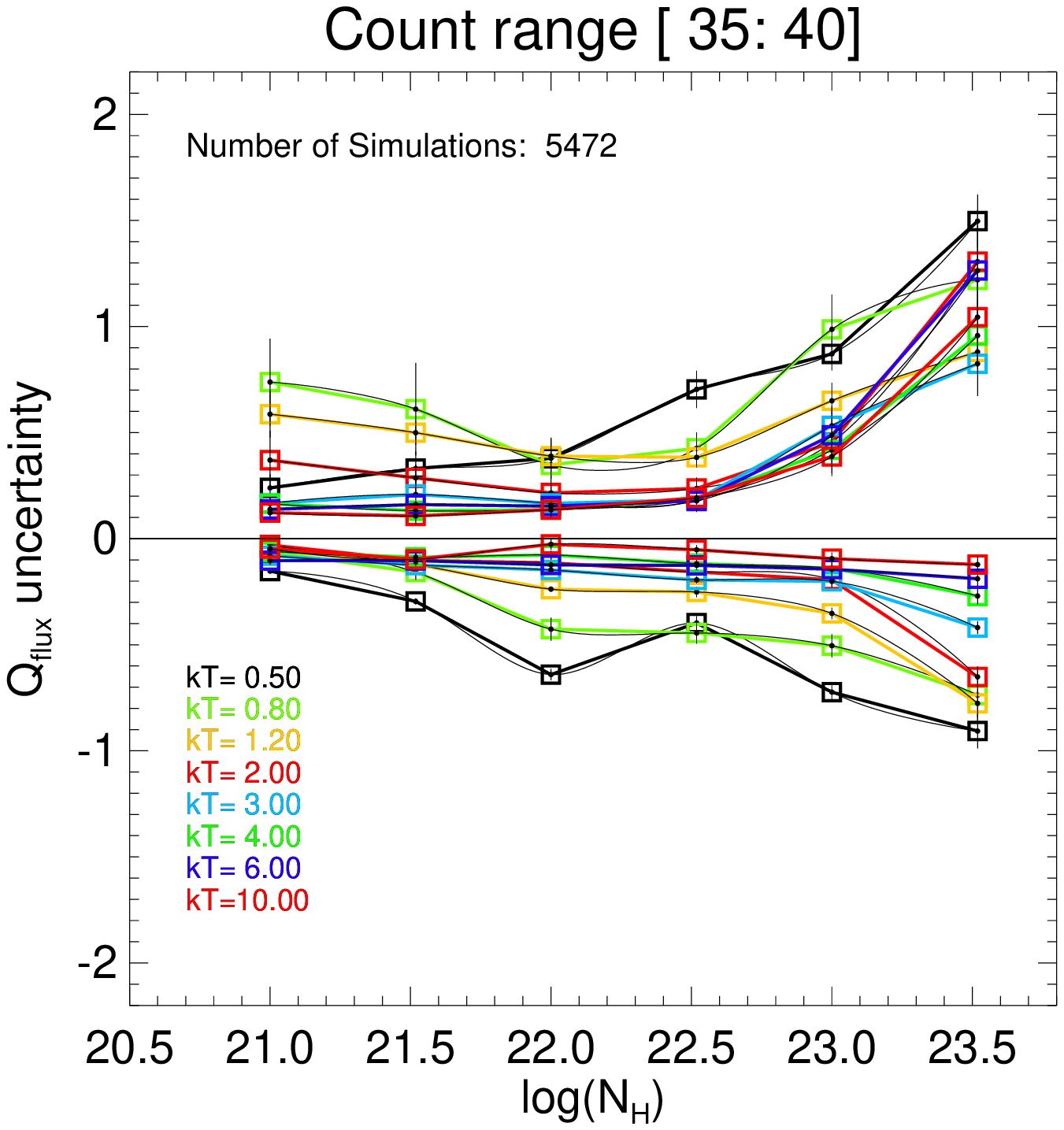}
\includegraphics[width=7.4cm,angle=0]{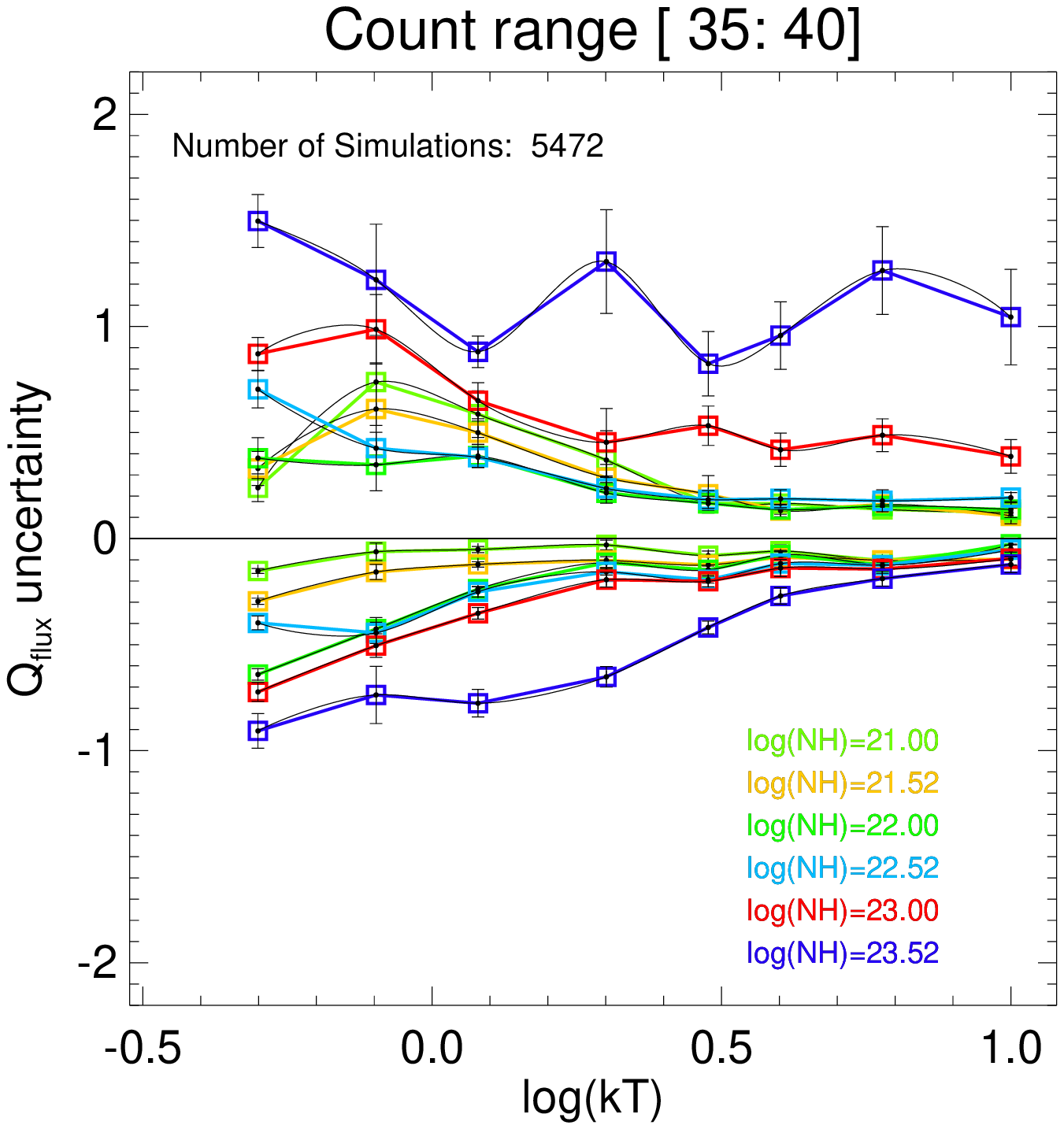}
\includegraphics[width=7cm,angle=0]{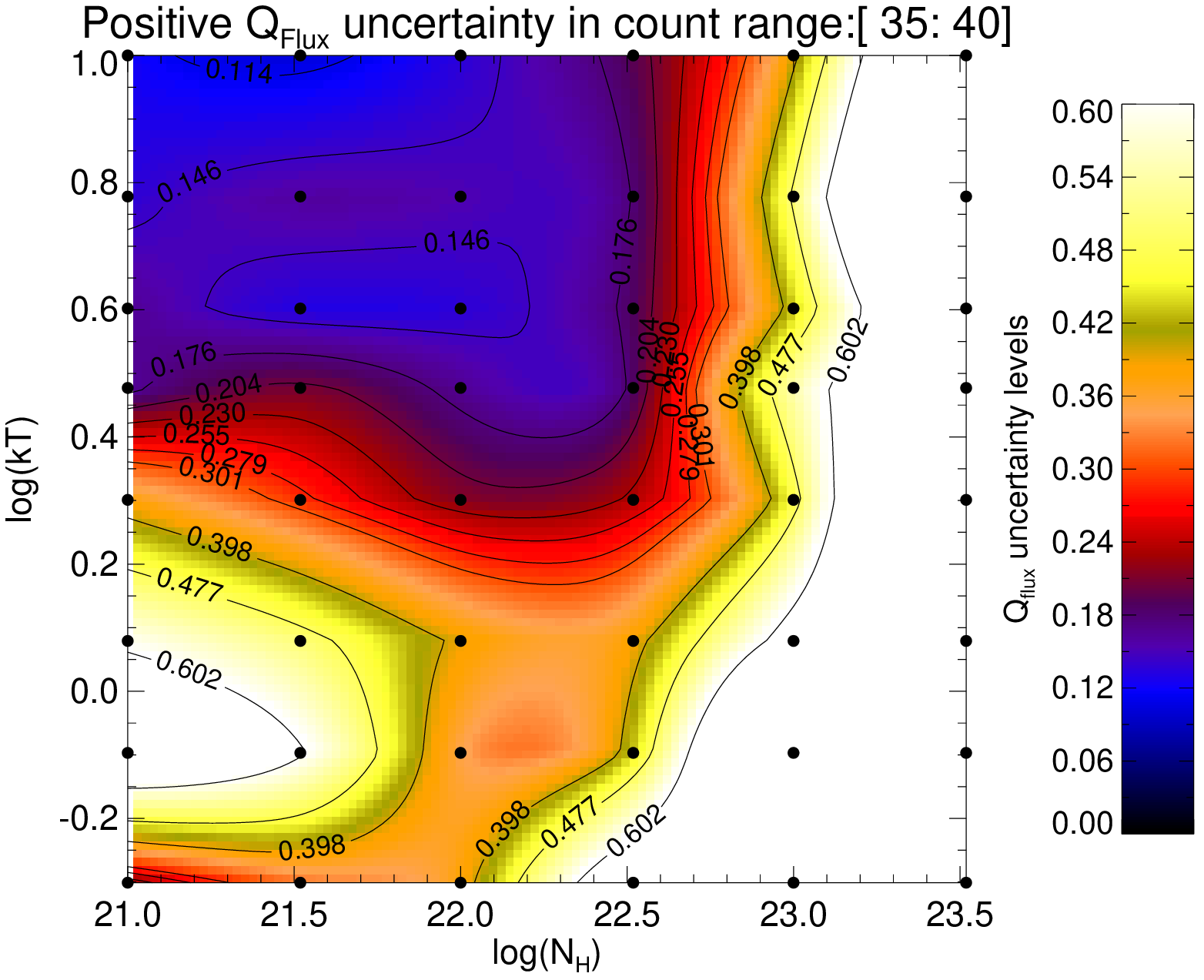}
\includegraphics[width=7cm,angle=0]{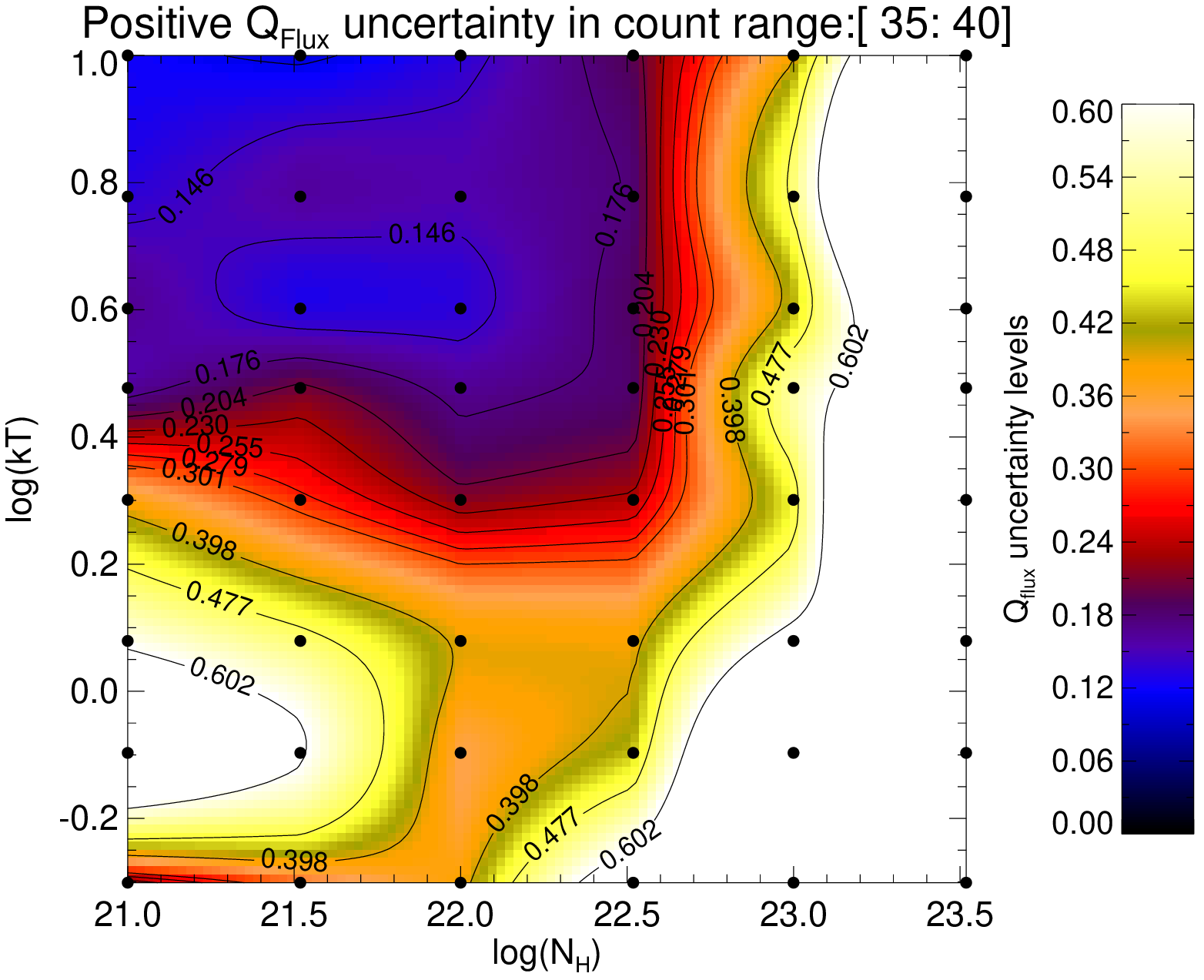}
\includegraphics[width=7cm,angle=0]{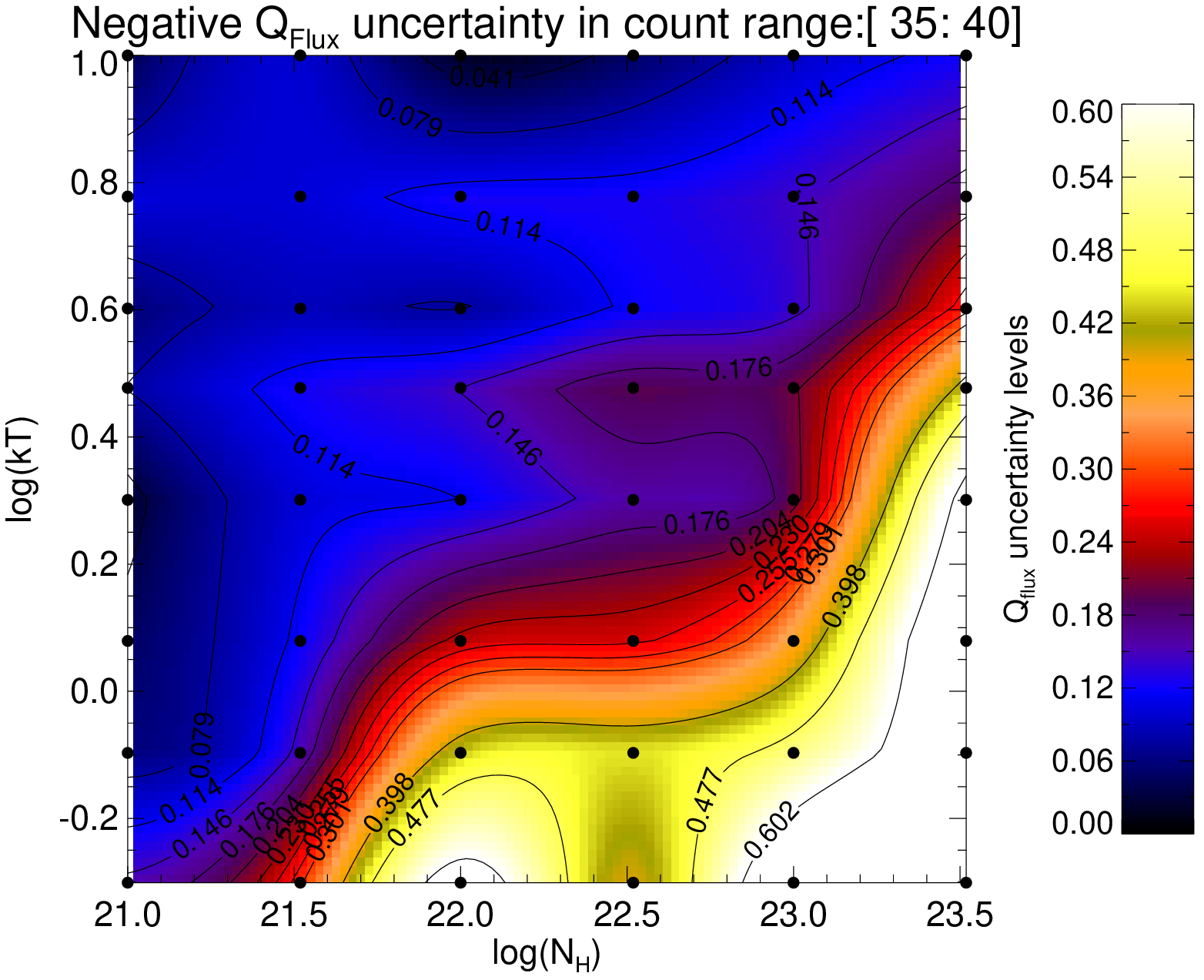}
\includegraphics[width=7cm,angle=0]{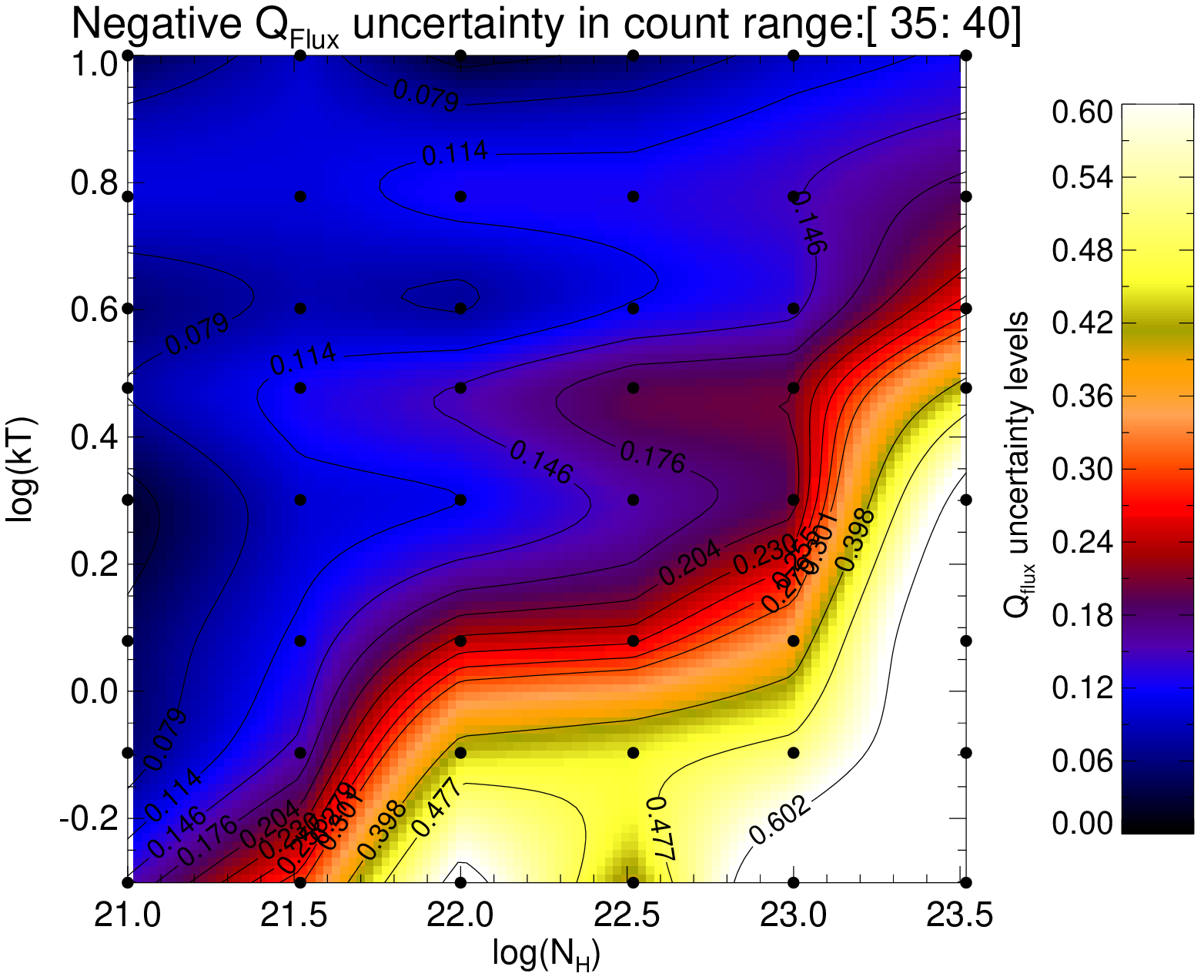}
\caption{Flux Q-uncertainty as a function of $N_{\rm H}$ and $kT$. 
The left column upper panel shows a set of eight different kT curves 
interpolated using a cubic spline function for MC simulated spectra in the range of 
35 to 40 photons. The Q$_{\rm flux}$ uncertainty is shown in left column as 2D bi-dimensional maps 
which was computed from the kT (y-axis) curves followed by a linear interpolation along the 
orthogonal ($N_{\rm H}$) direction (x-axis). The upper right plot shows the median 
Q$_{\rm flux}$ uncertainties (boxes) for the six different $N_{\rm H}$-curves also 
interpolated by a cubic spline function and for the same range of photon counts. 
The rest of right column figures show the Q$_{\rm flux}$ uncertainty as bi-dimensional map 
computed in an inverse interpolation order, i.e. along the kT direction. Note that for the 2D maps, 
both columns are essentially the same, with only very small differences, that suggest no interpolation artefacts 
were created in our numerical treatment of Q uncertaintities.}
\label{2Dgrid}
\end{figure*}

The outcome of our extensive set of MC simulations is a large collection
of best-fit spectral parameters from the fitting of simulated spectra.
We will focus on N$^{\rm fit}_{\rm H}$, kT$^{\rm fit}$/$\Gamma^{\rm
fit}$ (for thermal/non-thermal models), and, importantly, F$^{\rm
fit}_{\rm x}$, the absorption-corrected X-ray flux, and study how the
Q uncertainties on the determination of these parameters depend both on
photon statistics and on their values.  

In Figure~\ref{quantiles}, we show, for two model spectra, the runs of best-fit F$_{\rm
x}$, N$_{\rm H}$, and kT as a function of the number of source counts. Also shown are the 
$\pm 1\sigma$ quantiles (Q$^-_Y$ and Q$^+_Y$) computed for source net counts in the range 10--350.
The upper and lower panels refer to
a relatively hard and highly absorbed thermal source
($N_{\rm H}=10^{22}$\,cm$^{-2}$, $kT = 3.0$\,keV), and to a softer, less absorbed
source ($N_{\rm H}=10^{21}$\,cm$^{-2}$, $kT = 1.2$\,keV), respectively, illustrating how
model dependent our results are.  We note in particular difficulties in constraining the absorption for the latter, less absorbed, model, for which lower limits to the $N_{\rm H}$ parameter were often not restricted by the data for cases in which there were less than 100 net counts.  In these cases, the blue squares representing the lower quantile limits are set to the lower limit of the allowed range we specified in the XSPEC fitting procedure---a factor of 100 less than the input value.  This lack of sensitivity to low $N_{\rm H}$ values occurs because of the limited soft X-ray response of ACIS-I; this has been eroded since launch by the gradual build-up of a molecular contamination layer that attenuates low-energy X-rays \citep[e.g.][]{ODell2013}.  This limitation places a constraint on the range of validity of our results in the sense that uncertainties in $N_{\rm H}$ will be strictly accurate for $N_{\rm H}\ga 10^{21}$~cm$^2$.  For lower column densities both the $N_{\rm H}$ and unabsorbed flux uncertainties are subject to systematic error, the latter being affected because it is dependent on the former.  We return to this briefly in the verification of our method in Section~\ref{s:test_uncert}).

Another useful way to visualize our results is illustrated in
Fig.\,\ref{2Dgrid} (upper and lower left panels).  Here we show, for
models with total counts in a given range (in this case 35--40
counts), the flux uncertainties/quantiles as function of input kT for different input values of $N_{\rm H}$
(upper panel), and input $N_{\rm H}$ for different input values of kT (lower panel).
These figures illustrate that the dependence of uncertainties on model parameters 
is non-trivial and cannot easily be approximated analytically.  

The three-dimensional thermal and non-thermal grids of simulations also allow 
us to aspects of the results easily in two dimensions (three dimensions 
is also of course possible, but challenging to interpret). Also illustrated in the central 
and right-hand panels of Fig.\,\ref{2Dgrid} is an example of how contour maps of the two-dimensional 
dependence of the positive and negative 1$\sigma$ relative flux uncertainties depends on 
$N_{\rm H}$ and kT for the case of thermal spectra with 35--40 counts. To avoid 
discontinuous artefacts resulting from the limited sampling of our grid, these maps were 
produced by interpolating within the grids using a cubic spline function.  We tested this 
method by interpolating in both N$_{\rm H}$ for fixed kT values (top middle and right 
panels), and in kT for fixed  N$_{\rm H}$ values (bottom middle and right panels). The 
two approaches are in very good agreement, indicating that the log($N_{\rm H}$)--log(kT) 
space is well mapped, and that the approach is free from significant numerical instability 
or bias. The contour curves in Fig.\,\ref{2Dgrid}  were computed for logarithmic relative 
error values equal to log([1.001, 1.1, 1.2, 1.3, 1.4, 1.5, 1.6, 1.7, 1.8, 1.9, 2.0, 2.5, 3.0, 4.0]).

At this point, we are able to give a full description of the Q-uncertainty for any value 
of N$_{\rm H}$, kT or $\Gamma$, and X-ray flux, for a given discrete range of net counts 
(see Appendix) . However, as we mention in Section 2, the X-ray spectral simulations were 
performed by adopting a negligible background fraction in the spectra. 
Obviously, these results cannot be applied to those sources that suffer from higher background 
contributions.  For on-axis observations with {\it Chandra} ACIS-I, for which the PSF is very 
compact, background can often be very low compared with source counts. 
However, the PSF angular size increases with increasing off-axis 
angle\footnote{see http://cxc.harvard.edu/proposer/POG/html/chap4.html for details} and 
consequently the fraction of the photon events that comprise an extracted source spectrum 
that can be attributed to background also increases. The background contribution can be 
particularly significant for large off-axis angles ($\ga 3\arcmin$).  We address this problem in the next section.\\ 

\section{How background photons affect the Q uncertainty}
\label{s:bkgquncert}

\begin{figure*}[ht!]
\centering
\includegraphics[width=7.9cm,angle=0]{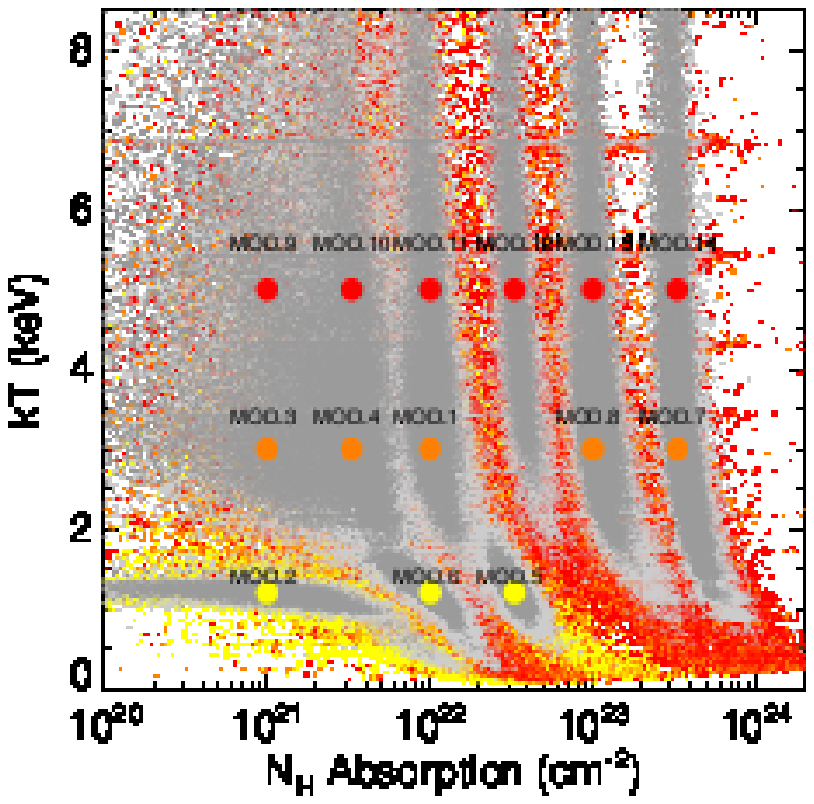}
\includegraphics[width=7.9cm,angle=0]{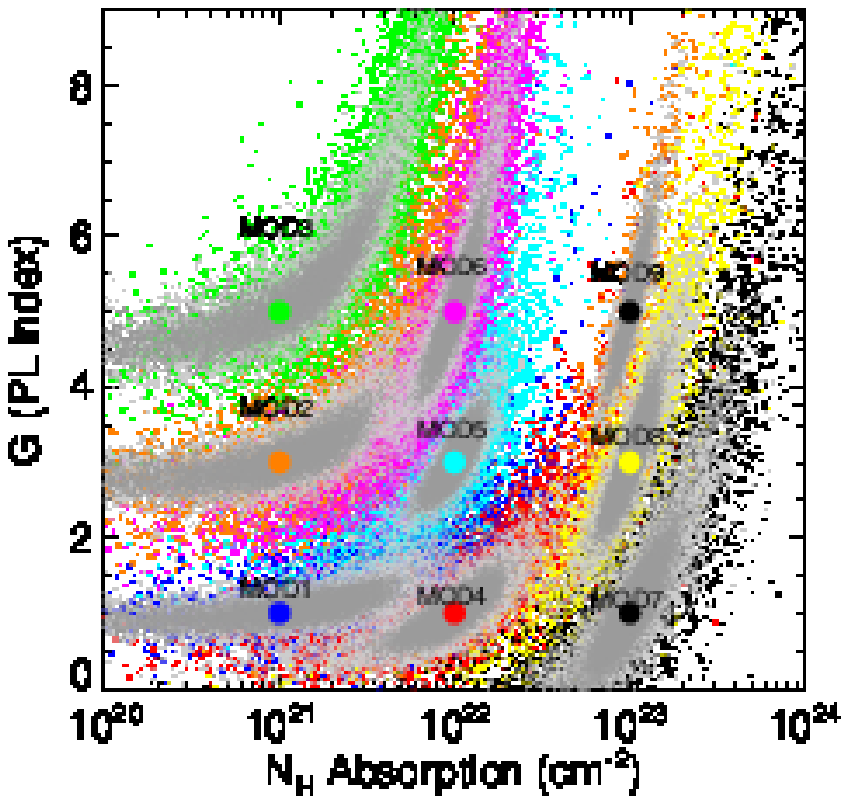}
\caption{Left: $N_{\rm H}$-kT space of solutions from $\sim$~500,000 MC simulations 
for different models, net counts, and background fractions. The different colors refer to 
the simulated thermal spectra at the fourteen different  $N_{\rm H}$-kT input values (big color circles). 
Small color dots are those that corresponds to MC simulations for 20 and 30 net counts in the spectra.
Light grey and darker grey dots show the MC simulations for net counts of 100 and 300 photons 
in the spectra respectively.  The clustering of points giving rise to a small degree of horizontal 
striping along the direction of the kT axis corresponds to slightly biased spectral fit solutions 
that mostly occur for MC simulations with 20 and 30 net counts (see text). Right:  A similar illustration of a 
total of $\sim$~192000 non-thermal MC simulations for different $N_{\rm H}$-$\Gamma$ models 
and background fractions.}
\label{plane_model}
\end{figure*}

\begin{figure*}[ht!]
\centering
\includegraphics[width=7.9cm,angle=0]{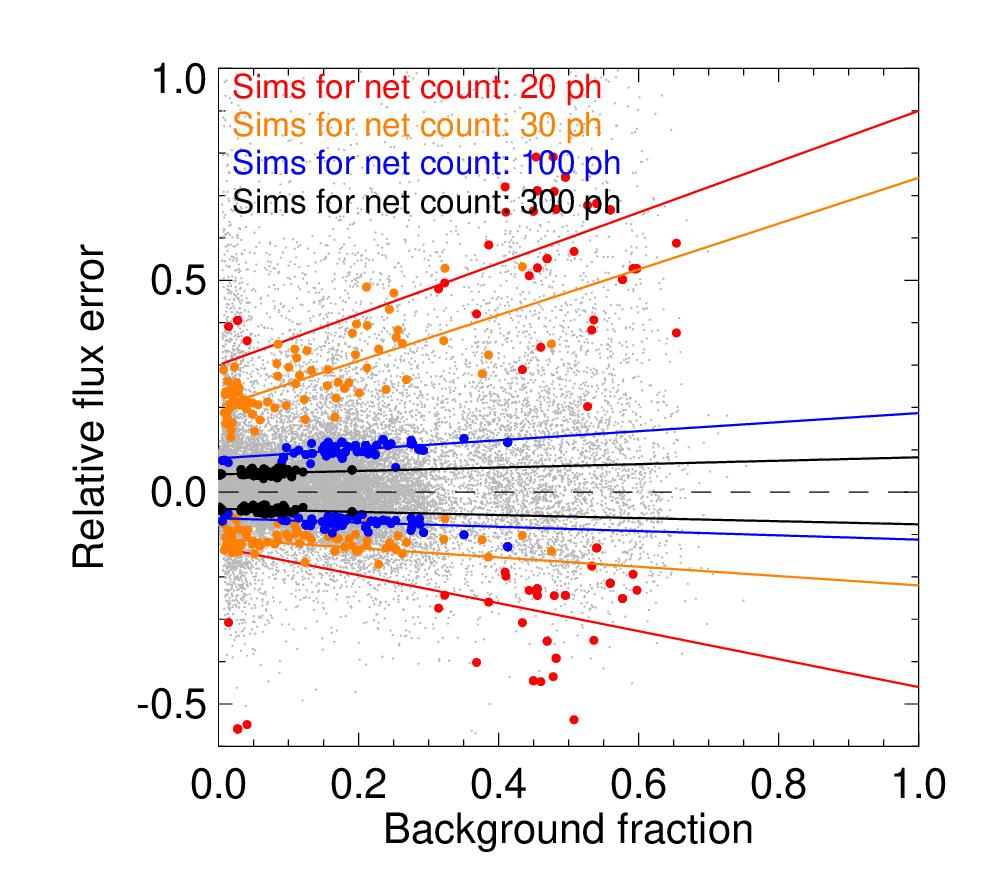}
\includegraphics[width=7.9cm,angle=0]{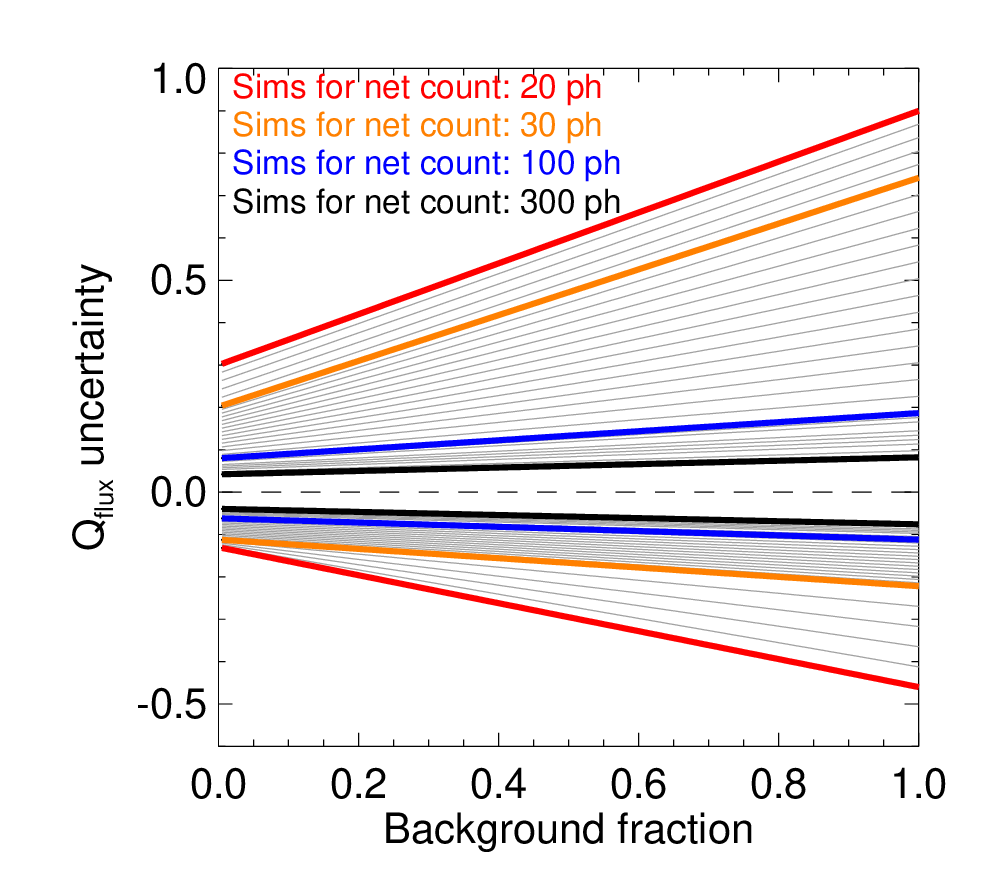}
\caption{The left panel illustrate curves fitted to the relative flux 
error as a function of the background fraction for the thermal model 1 
[N$_{\rm H}$=1.0$\times$10$^{22}$ , kT=3.0] according to MC 
simulations containing 20 (red), 30 (orange), 100 (blue) and 300 (black) net 
photons in the spectra, respectively (see text).  Solid color large points in the 
panel are the Quantile uncertainties for flux. 
The relationship between the flux uncertainty and background fraction is clearly quite linear.  
The right figure illustrates the synthesis of the fits to the data of the left panel. 
The whole set of interpolated curves for the rest of the model parameters investigated 
and for each of fourteen models were computed but not shown here. The same numerical treatment  
was performed for non-thermal models and for all spectral parameters.}
\label{qflux_model}
\end{figure*}

The results presented in Sections~ 2 and 3 corresponds to the case of sources with negligible background contamination, i.e., a 
background fraction less than 0.01.  As noted above, this is generally not the case for {\it Chandra}, or many other X-ray 
observations, for which background events can be a significant or even dominant fraction of the extracted signal.
High background contributions can potentially mask or bias the true nature of source X-ray spectra, make spectra 
noisy and compromise their interpretation and the ability to constrain spectral parameters in astrophysical modelling. In 
this section, we study this effect by means of further extensive MC simulations of background-affected spectra.

As before, the main goal is to estimate the dependence of the uncertainty of fitted spectral parameters on the intrinsic
X-ray emission source model and the net photon counts of the spectra.  We constructed a set of simulations based on 
eight different thermal and non-thermal X-ray emission models, simulated for different numbers of source and background 
counts. The background fraction of the simulations ranged between 0.001 to the limiting case of 1 (corresponding 
to a spectrum comprised of 100\%\ background photons). We adopted RMF, ARF and background (BKG) files that correspond to a set of 
100 X-ray sources randomly distributed over the field-of-view of the ACIS-I {\it Chandra} observation ID~4511 of the 
Cygnus OB2 association (see \citealt{Albacete2007} for further details of the observation).

Owing to the additional dimension of background fraction, we covered a more limited segment of the $N_{\rm H}$--kT plane 
than in the zero background grid employed in Sect.~\ref{sect:xsims}. Thermal and non-thermal models were 
computed for a set of temperatures and/or $\Gamma$ index values that are listed in Table\,\ref{models} and
also illustrated in Figure~\ref{plane_model}. This amounted to a set of 14$\times$100 (thermal case) simulations 
for background fractions from 0.001 to 1, and for total events numbering 20, 30, 100 and 300 photons. 

In the left of Figure\,\ref{plane_model}, we show the the distribution of fit 
solutions for each model and for the different net counts in the spectra and background contamination fractions.  At low absorption, 
the retrieved $N_{\rm H}$ uncertainty is greater than that in temperature, while the opposite behavior occurs when $N_{\rm H}$ becomes 
large, leading to a poor determination of temperature. This behavior arises simply because of the way the spectral ``lever arm'' acts: at 
low $N_{\rm H}$, there is little energy range over which the spectrum is significantly affected by absorption and purchase on $N_{\rm H}$ 
is weak---as in the case of unconstrained $N_{\rm H}$ lower limits noted in Section~\ref{s:analuncert} above; conversely, there is a large energy range available to discern the spectral shape characteristics that are affected by temperature.  
At high $N_{\rm H}$, the opposite is true.  Furthermore, uncertainty in temperature grows at higher temperatures because of the limited 
broad-band energy range available for our spectral fit.  The spectrum tends to flatten out within the 8.0~keV bound we adopt; at higher 
energies where the temperature discrimination power largely lies for the hotter models, the ACIS-I effective area declines rapidly.   

Closer inspection of the left panel in Figure\,\ref{plane_model} reveals some striation in the distributions of best-fit parameters, indicating that certain quantized values of temperature are slightly preferred over others.  These artifacts result from the fits to spectra with low numbers of counts ($\leq$ 30 photons) and the preferred temperatures are those of the plasma radiative loss model temperature grid.  They occur when the uncertainties on the spectral fit are large, much larger than the grid quantization.  This apparent fit bias only affects a very small fraction of the total and does not impose any bias on fit quantiles because the quantization effect is on a scale smaller than parameter estimation uncertainties. 

A similar procedure to that for thermal models was followed to compute non-thermal models. A $3\times$3 grid of power-law indices ($\Gamma=1,3,5$) 
and absorbing column densities ($N_{\rm H}=10^{21},10^{22},10^{23}$) was adopted, amounting to  9$\times$100 MC 
non-thermal simulations for background fractions in the range 0.001 to 1, and for 20, 30, 100 and 300 photons 
(see the right panel of Figure~\ref{plane_model}).  The striation effect discernible for the thermal model fits does not occur for non-thermal models because the spectral model itself is analytic and not stored in quantized, tabulated form.

\begin{table}
\caption{X-ray emission models used to investigate the effect of background contamination fraction
in the determination of Q uncertainties of spectral parameters from simulated spectra.}
\begin{center}
\begin{tabular}{llllllll}
\hline
Model    	& \multicolumn{3}{c}{Thermal}  		&& \multicolumn{3}{c}{Non-thermal} 	\\ 
\cline{2-4}
\cline{6-8}
\#		& N$_{\rm H}$	&  kT  	& N$_{\rm sims}$&& N$_{\rm H}$ 			&  $\Gamma$ 	& N$_{\rm sims}$\\
		&(cm$^{-2}$)	& (keV)	&			   && (cm$^{-2}$) 	&			&			\\
\hline
1	&	1.0$\times$10$^{22}$&	3.0	& 32017&&	10$^{21}$	&	1.0 	&26031\\
2	&	0.1$\times$10$^{22}$&	1.2	& 35779&&	10$^{21}$	&	3.0 	&29467\\
3	&	0.1$\times$10$^{22}$&	3.0	& 31413&&	10$^{21}$	&	5.0 	&29304\\
4	&	0.33$\times$10$^{22}$&	3.0	& 49918&&	10$^{22}$	&	1.0 	&21967\\
5	&	3.3$\times$10$^{22}$&	1.2	 &50159&&	10$^{22}$	&	3.0 	&15269\\
6	&	1.0$\times$10$^{22}$&	1.2	 &48968&&	10$^{22}$	&	5.0 	&23032\\
7	&	33.0$\times$10$^{22}$&	3.0	 &46859&&	10$^{23}$	&	1.0 	&22730\\
8	&	10.0$\times$10$^{22}$&	3.0	 &31762&&	10$^{23}$	&	3.0 	&18776\\
9	&	0.1$\times$10$^{22}$ &	5.0	 &45730&&	10$^{23}$	&	5.0 	&7172\\
10	&	0.33$\times$10$^{22}$ &	5.0	 &44960&&	$--$	&	$--$ 		&\\
11	&	1.0$\times$10$^{22}$ &	5.0	 &41042&&	$--$	&	$--$ 		&\\
12	&	3.3$\times$10$^{22}$ &	5.0	 &34020&&	$--$	&	$--$		&\\
13	&	10.0$\times$10$^{22}$ &	5.0	 &45116&&	$--$	&	$--$		&\\
14	&	33.0$\times$10$^{22}$ &	5.0	 &49918&&	$--$	&	$--$		&\\
\hline
\end{tabular}
\end{center}
\label{models}
\end{table}

In order to get a quantitative estimate of the uncertainty in best-fit spectral parameters, we compute their relative errors 
and respective quantiles as a function of the background fraction. Uncertainties correlate linearly with the background
 fraction of the simulated spectra, being well described by a function of the form
\begin{equation}
Err({\rm bkg\_frac,cnt}) =  A({\rm cnt})\times{\rm bkg\_frac} + {\rm B({\rm cnt})}.
\label{e:err}
\end{equation}

The coefficients of such a relationship are not the same for all spectral models even if they have the same net numbers 
of counts in the simulated  spectra. They depend on the emission characteristics of the source, as well as the 
signal-to-noise ratio of the simulations. We evaluated the linear dependence of all the fourteen thermal, and nine non-thermal, 
models and computed the explicit dependence of Q on bkg\_frac. In Table\,\ref{models}, we list the parameters of the 
background affected spectral models to map $N_{\rm H}$-kT or $N_{\rm H}$-$\Gamma$ planes, and the total of 
simulations performed for each one.
Of the coefficients A (the slope) and B (the intercept), the latter agrees with the 1$\sigma$ quantile error estimation 
for MC simulations of spectra in which the background fraction is negligible or zero (bkg\_frac~$<$~0.01). We thus 
can rewrite Equation~\ref{e:err} as
\begin{equation}
Err({\rm bkg\_frac,cnt}) =  A({\rm cnt})\times{\rm bkg\_frac}. + {\rm Q_0({\rm cnt})}.
\label{e:errrw}
\end{equation}

In Table~\ref{t:errorcoef} of the Appendix we list explicit $A({\rm cnt})$ coefficients computed from the uncertainty 
curves for 20, 30, 100 and 300 photons in background-affected MC simulated spectra. Such 
coefficients represent the change of the uncertainty estimation Q$_{\rm 0}$ due to the influence of 
background photons in the net counts of the X-ray spectra. The most straightforward way to get 
the adequate $A$ coefficient for a given X-ray source is by choosing the most representative model 
for the true source X-ray spectrum.
 
We can linearly interpolate within the grid of results to obtain bi-dimensional planes of the 
uncertainties in any given parameter as a function of background fraction for a given number 
of total counts between 20 and 300.  As an example, in Figure\,\ref{qflux_model} we show how 
the uncertainty in X-ray flux increases with the background fraction, and how it is more tightly 
constrained as the net counts in the spectra increases. The thermal "model 1"  ($N_{\rm H}=
10^{22}$ and kT=3.0 keV; see the Figure~\ref{qflux_model}) was chosen as the example for 
this figure, and corresponds to the most representative model of the sources in the Cyg\,OB2 
Association (which are mostly T~Tauri stars; \citealt{Albacete2007}, Flaccomio et al., 2016, this issue; 
Drake et al., 2016, this issue).  We discuss the Cyg~OB2 case in more detail in Section~\ref{s:test_uncert} below.\\

\begin{figure}[ht]
\centering
\includegraphics[width=8.8cm,angle=0]{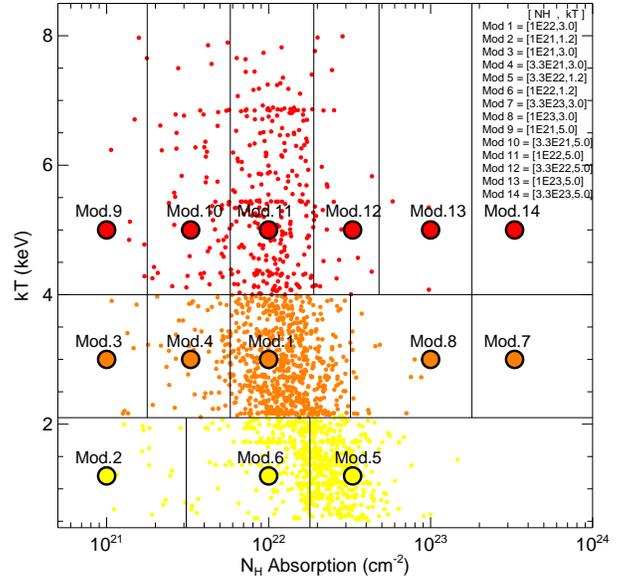}
\caption{The kT-$N_{\rm H}$ plane for parameter values obtained from X-ray spectral fits to Cyg\,OB2 
in sources having a signal strength in the range of 20 to 350 net photons. 
The nine colored models are those used to compute the influence of the different background 
fractions in the determination of uncertainty for a range of spectral parameters.}
\label{nh-kt_plane}
\end{figure}

\section{Testing uncertainties of X-ray model parameters for faint Cyg\,OB2 sources}
\label{s:test_uncert}

\begin{figure*}
\includegraphics[width=12.7cm,angle=0]{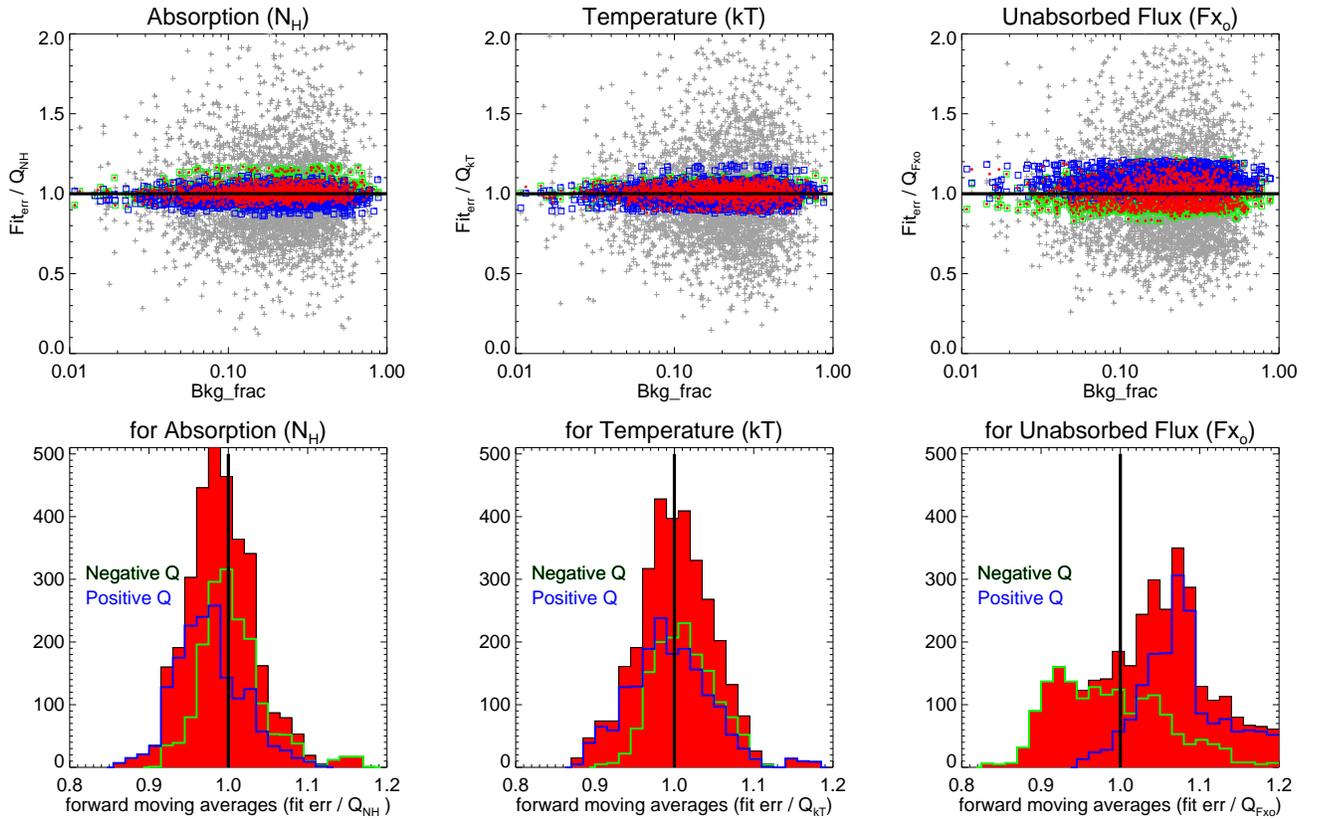}
\caption{
Upper panel: The ratios of individual fit errors from non-interactive spectral 
fitting to the Q uncertainty computed from MC simulations (i.e. Fit\_err/Q), for each of the sources that were included in the 
comparison analysis (see section~\ref{s:comp_err_fit_q}).  Error ratios were computed for the three main parameters, NH, 
kT and unabsorbed flux. Grey crosses are the individual source Fit\_err/Q ratios for combined positive and negative uncertainties, 
with errors representing $1\sigma$ uncertainties. 
Small green and blue dots correspond to the forward moving averages (see text for description)
as a function of the increasing background fraction for positive and negative $1\sigma$~quantiles, respectively. 
Small red dots were computed in the same way but assuming both positive and negative quantile distributions 
as a single one.
Lower panels are simply histograms taken on the y-axis (mean Fit\_err/Q) of the corresponding set of 
colored points in the upper panels.}
\label{Qvserr}
\end{figure*}

One main motivation of this work is to estimate the uncertainties on luminosities and spectral 
parameters expected for sources yet to be observed, based on the anticipated ACIS-I {\it Chandra} 
X-ray spectrum. However, another use is to cross-check, or even calculate, the uncertainty obtained 
from the fitting process and to give rough estimates in those cases in which the calculation of 
uncertainties fails formally (e.g.\ when the {\sc error} command in XSPEC fails due to numerical issues). 
As a first application and verification of our method, we use real data and spectra that have been extracted as part 
of the Cygnus~OB2 Chandra Legacy Survey (Drake et al. this issue). In this context, our aim is: i) to 
give spectral fit constraints from our quantile method for Cyg\,OB2 sources; ii) to compare the 
uncertainties in the X-ray spectral parameters for faint sources with those results obtained from 
non-interactive spectral fits presented in Flaccomio et al. (this issue).

Following results of the analysis presented by \cite{Wright2014}, a large fraction of the 
Cyg~OB2 sources have been detected at large off-axis angles with ACIS-I. As we 
have noted above, the PSF degradation of the ACIS-I detector with increasing off-axis angle means the relative contribution 
of background photons to source events can be relatively large, which can affect the way in which spectral 
fitting converges as well as the uncertainty in the computed spectral solution. 
In fact, of the 7924 Cyg~OB2 sources, just 10 have background fractions less than $\sim 0.01$.  The median of 
the distribution of background fractions is $\sim$ 0.36, with some background 
fractions being as high as 0.9. Thus, if we want to get an unbiased $1\sigma$ Quantile error estimation, 
it is strictly necessary to include the background dependency while accounting 
for the different X-ray emission characteristics of the sources.  We employ a set of nine different 
optically-thin thermal plasma models that in the kT-$N_{\rm H}$ plane represent the range in the 
uncertainty of parameters and flux as a function of the different background fractions (see 
Fig\,\ref{nh-kt_plane}). This procedure allows us to get a much more precise estimate of the 
influence of the background on parameter uncertainties  than we would be able to from the use of 
a single average model. 

The complete catalog of Cyg\,OB2 sources in the survey comprises a total of 7924 X-ray objects 
\citep{Wright2014}.  Of these, 4975 have less than 20 net photons, i.e.\ about 62\% of 
the whole source sample.  Flaccomio et al (this issue, 2016) have performed spectral fitting of X-ray spectra 
for sources with more than 20 and less than 350 counts, and if we restrict the sample to this 
range we get a total of 2755 sources for which X-ray spectral fits could give an 
error estimation for the model parameters and flux.   We further limit ourselves to X-ray sources 
that were classified as members of the Association (Kashyap et al, 2016, this issue) and whose X-ray 
spectral fitted parameters are in the range of $N_{\rm H}$ [$1\times 10^{21}$--$3.3\times 10^{23}$] and 
kT = [0.5-8] keV. This subsample contains 1718 sources, and corresponds to 20\% of the entire sample
Cyg~OB2 members.\\

\subsection{Comparing errors from spectral fits with Quantile estimates}
\label{s:comp_err_fit_q}

In this subsection we study the goodness of our method in the determination of uncertainties
of the X-ray parameters computed from spectral fits. To do that we need to accounts for two 
approaches: i) to compute the fit error of a given parameter (fit\_err) via the XSPEC error command, 
which consist in the 1 $\sigma$ error computed from the covariance matrix in the spectral fits of Cyg~OB2 
stars (see Flaccomio et al. this issue); ii) to determinate the respective Q-uncertainties 
computed from our 2D Quantile maps corrected by the influence
of the individual X-ray background fraction of the Cyg OB2 sources. The ratio between 
these two quantities is show as grey crosses in the Figure\,\ref{Qvserr} (upper panel).

In order to give a robust quantitative estimates of the reliability of our method, we made use of
a statistical estimator known as the Forward Moving Averages (FMA), which can be computed from the 
cumulative unweighted averages of a given set of sorted sample data points. 
We sort all the fit\_err/Q values in ascending order
of background fraction. The FMA starts from the first fit\_err/Q value until the next recomputed current 
($n+1$) value, to the last adopted data point of our sample. 
Further details can be found from the help of the 
{\sc ts\_smooth}\footnote{http://www.exelisvis.com/docs/TS\_SMOOTH.html.} IDL task. 
We use this technique on parameters $N_{\rm H}$, kT, and unabsorbed X-ray flux, to compute the goodness 
quantile error prediction with respect to the errors computed from X-ray spectral fits. 
As is shown in Figure\,\ref{Qvserr}, the majority of distributed FMA 
values are less than 10--12 \%, even for higher background fractions.

While the combined positive and negative uncertainties from our approach are in excellent agreement 
with uncertainties from spectral fits to observed data, we note that the one-sided uncertainties do show 
some systematic deviations, albeit at levels of only $10$~\%\ or so.  This is revealed by the spread in 
the colored dots in the top panels of Figure\,\ref{Qvserr}, and in the separation of the distributions in 
the lower panel.  There are good reasons why the uncertainties should {\em not} be in perfect agreement.  
Fits to observed data use only an approximation to the true instrument response---represented by the 
ARF and RMF files noted in Section~\ref{sect:xsims}---and also assume a spectral model that will 
diverge from the real emission spectrum of a plasma at some level. In contrast, the simulations are fit 
with exactly the same models used to generated them, such that the difference between the retrieved 
and input spectral parameters results from Poisson noise.  In the regime of a large number of photon 
counts and relatively low background, these systematic differences will be more important.   

Systematic errors in fitted spectral parameters due to "imperfect" instrument response functions and 
plasma model spectra will lead to systematic errors in the derived quantile because the point of 
interpolation in the 2D quantile maps will be offset from where it should ideally be. In regions of these 
maps that are relatively flat, this interpolation error will be small. However, there are some regions of 
these quantile planes that have larger gradients, and in these regions the systematic error in the 
derived quantile can be larger.  

At the low signal-to-noise end, we have also encountered in Section~\ref{s:analuncert} the low hydrogen 
column density cases in which the spectral fit simulations are occasionally unable to constrain the lower 
bound $N_{\rm H}$ fit errors.  Indeed, the discrepancy in the two-sided errors is largest for the unabsorbed 
flux, which in the bandpass of interest here is highly dependent on the uncertainty in $N_{\rm H}$.  The 
two-sided errors are somewhat more discrepant for unabsorbed flux than for the $N_{\rm H}$ absorption or
kT plasma temperature, which are in excellent agreement. {\it Chandra} ACIS-I spectra have good observational 
leverage on plasma temperature for typical Cyg~OB2 sources. However, modeling degeneracies 
between $N_{\rm H}$ and kT are less well constrained for less absorbed stars with low signal-to-noise 
spectra, potentially leading to inaccuracy in the error estimate for the unabsorbed flux.
In the light of this discussion, we conclude that in general the error estimation from spectral fitting 
is in very good agreement with our quantile estimates.  Our method then represents a reliable means for 
error estimation in X-ray spectral parameters whenever the model fit to the X-ray data cannot provide it, 
as well as for observation planning in which data are not yet in-hand.  The method can also be applied to 
promptly estimate the errors in spectral parameters for previously studied X-ray source populations when 
uncertainties where not published or are otherwise unavailable.\\

\section{Summary}

We present a new method for the estimation of errors in X-ray spectral parameters and fluxes
of faint sources detected by ACIS-I onboard {\it Chandra}. X-ray sources in the range of 10 to 
350 net photons are generally not well-suited to quantitative spectral analysis. Simple X-ray 
spectral modeling of such weak sources can suffer from ill-constrained solutions and mismatches 
between true and modeled parameters, combined with unrealistic estimation of the true uncertainties.

Here, we have described an original treatment that resolves these problem by means of an extensive 
set of MC simulations of faint X-ray spectra.  As representative cases, we adopted absorbed optically-thin 
thermal plasma models and power-law continua that are commonly employed to interpret a wide range of 
astrophysical sources of X-rays. We fitted simulated spectra for a wide range of absorption, plasma 
temperature and power-law index. We studied the relative error distributions of retrieved parameters and 
fluxes and describe fits in terms of 1$\sigma$ quantiles. We computed parametric curves of quantiles as a 
function of given input model parameters and interpolated within these to compute bi-dimensional maps.
Such a set of maps provides an estimation of the true error in spectral parameters whenever it is not 
possible to determine these adequately from a direct spectral fit.

Our method makes it possible to improve statistical studies of objects in which a large fraction have a low 
number of counts.  As an example, we have applied this method to sources in the {\it Chandra Cygnus OB2 Legacy 
Survey} and the properties of faint X-rays sources that mostly correspond to young low-mass stars.  The uncertainties estimated using the quantile method presented here and from spectral fitting applied to the observations are in excellent agreement, and generally at a level of 10~\% or better.   
  
Finally, our method and data enable the computation of the exposure times required for surveys and 
individual pointed observations to reach a given uncertainty in derived spectral parameters.  The results 
of this work will be made publicly available in an on-line tool in the near future.

\acknowledgments

We thank the referee Dr. Maurice Leutenegger for many and very helpfully suggestions to our article.
JFAC is a researcher of CONICET and the University of Rio Negro (UNRN) and 
acknowledges support from grant CONICET in the 
context of an External post-doctoral fellowship at the Osservatorio Astronomico di Palermo (OAPA). 
EF acknowledge support from the INAF and to the OAPA.
JJD and VK were supported by NASA contract NAS8-03060 to the {\it Chandra X-ray Center} (CXC) 
and thank the director, B.~Wilkes, and the science team for continuing support and advice. 

\bibliographystyle{aa}
\bibliography{yaReferences}

\begin{thebibliography}{12}
\expandafter\ifx\csname natexlab\endcsname\relax\def\natexlab#1{#1}\fi

\bibitem[{{Albacete Colombo} {et~al.}(2007){Albacete Colombo}, {Flaccomio},
  {Micela}, {Sciortino}, \& {Damiani}}]{Albacete2007}
{Albacete Colombo}, J.~F., {Flaccomio}, E., {Micela}, G., {Sciortino}, S., \&
  {Damiani}, F. 2007, \aap, 464, 211

\bibitem[{{Anders} \& {Grevesse}(1989)}]{Anders1989}
{Anders}, E. \& {Grevesse}, N. 1989, \gca, 53, 197

\bibitem[{{Arnaud}(1996)}]{Arnaud1996}
{Arnaud}, K.~A. 1996, in Astronomical Society of the Pacific Conference Series,
  Vol. 101, Astronomical Data Analysis Software and Systems V, ed. G.~H.
  {Jacoby} \& J.~{Barnes}, 17

\bibitem[{{Broos} {et~al.}(2010){Broos}, {Townsley}, {Feigelson}, {Getman},
  {Bauer}, \& {Garmire}}]{Broos2010}
{Broos}, P.~S., {Townsley}, L.~K., {Feigelson}, E.~D., {et~al.} 2010, \apj,
  714, 1582

\bibitem[{{Cash}(1979)}]{Cash1979}
{Cash}, W. 1979, \apj, 228, 939

\bibitem[{{Guarcello} {et~al.}(2015){Guarcello}, {Drake}, {Wright}, {Naylor},
  {Flaccomio}, {Kashyap}, \& {Garcia-Alvarez}}]{Guarcello2015}
{Guarcello}, M.~G., {Drake}, J.~J., {Wright}, N.~J., {et~al.} 2015, ArXiv
  e-prints

\bibitem[{{Hong} {et~al.}(2004){Hong}, {Schlegel}, \& {Grindlay}}]{Hong2004}
{Hong}, J., {Schlegel}, E.~M., \& {Grindlay}, J.~E. 2004, \apj, 614, 508

\bibitem[{{Maggio} {et~al.}(1995){Maggio}, {Sciortino}, {Collura}, \&
  {Harnden}}]{Maggio1995}
{Maggio}, A., {Sciortino}, S., {Collura}, A., \& {Harnden}, Jr., F.~R. 1995,
  \aaps, 110, 573

\bibitem[{{O'Dell} {et~al.}(2013){O'Dell}, {Swartz}, {Tice}, {Plucinsky},
  {Grant}, {Marshall}, {Vikhlinin}, \& {Tennant}}]{ODell2013}
{O'Dell}, S.~L., {Swartz}, D.~A., {Tice}, N.~W., {et~al.} 2013, in Society of
  Photo-Optical Instrumentation Engineers (SPIE) Conference Series, Vol. 8859,
  Society of Photo-Optical Instrumentation Engineers (SPIE) Conference Series,
  0

\bibitem[{{Smith} {et~al.}(2001){Smith}, {Brickhouse}, {Liedahl}, \&
  {Raymond}}]{Smith2001}
{Smith}, R.~K., {Brickhouse}, N.~S., {Liedahl}, D.~A., \& {Raymond}, J.~C.
  2001, \apjl, 556, L91

\bibitem[{{Wright} {et~al.}(2014){Wright}, {Drake}, {Guarcello}, {Aldcroft},
  {Kashyap}, {Damiani}, {DePasquale}, \& {Fruscione}}]{Wright2014}
{Wright}, N.~J., {Drake}, J.~J., {Guarcello}, M.~G., {et~al.} 2014, ArXiv
  e-prints

\bibitem[{{Wright} {et~al.}(2015){Wright}, {Drake}, {Guarcello}, {Kashyap}, \&
  {Zezas}}]{Wright2015}
{Wright}, N.~J., {Drake}, J.~J., {Guarcello}, M.~G., {Kashyap}, V.~L., \&
  {Zezas}, A. 2015, ArXiv e-prints

\end{thebibliography}

\appendix

Here we present bi-dimensional maps of the Q uncertainty for 
flux and spectral parameters of thermal and non-thermal models.
As examples, we show 2D quantile maps restricted to 90 to 100 
X-ray photons in the simulated spectra. See the captions for a 
description of individual panels. For each case, movies 
covering all ranges of parameters investigated as a function of the X-ray 
net counts in the spectra are presented as on line material.

\begin{figure*}
\centering
\includegraphics[width=11cm,angle=0]{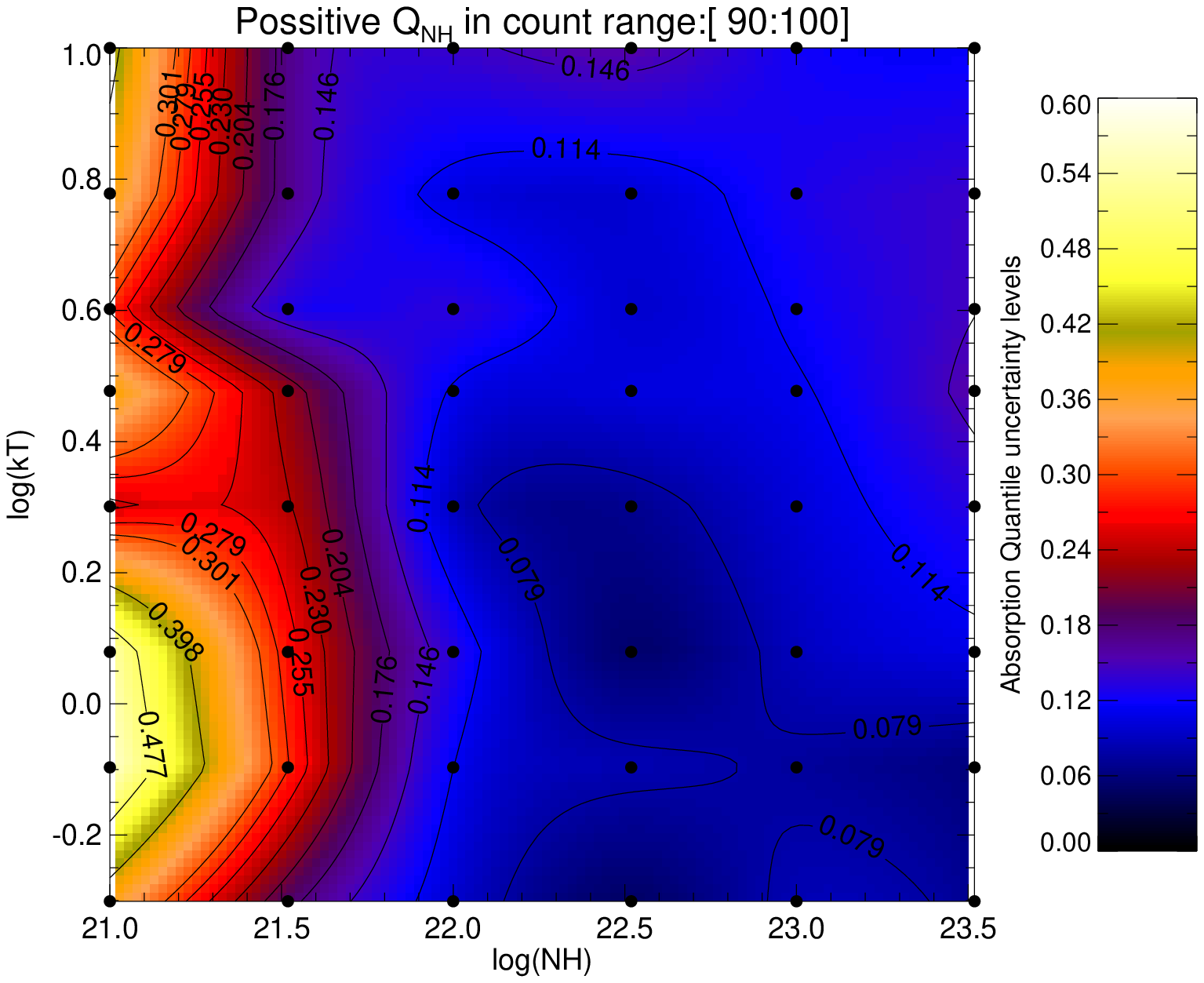}
\includegraphics[width=11cm,angle=0]{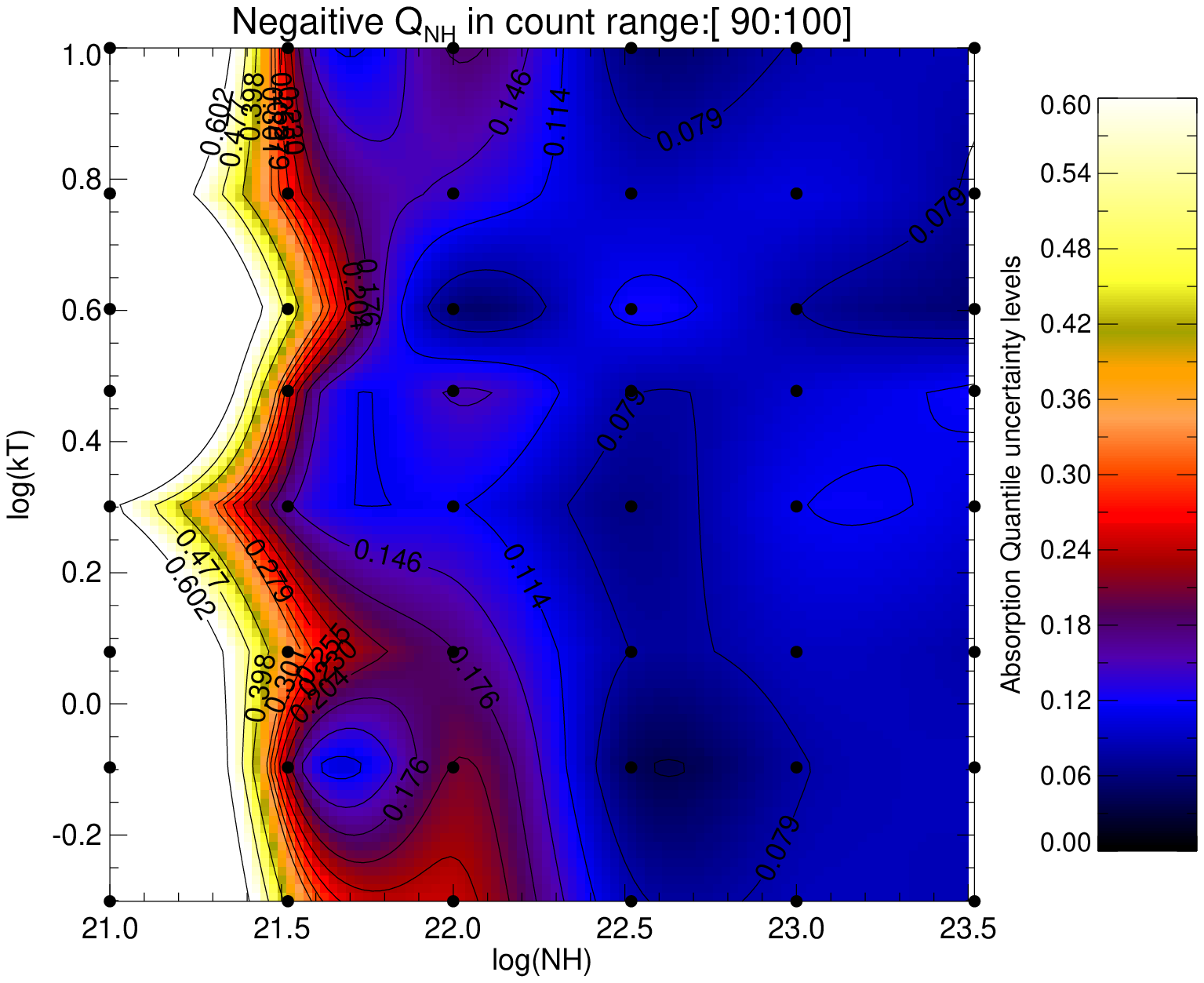}\\
\caption{\small \it 2D-N$_{\rm H}$ example Quantile maps from simulations of thermal models with 90 to 100 net photon counts.  Left and 
right columns refer to positive and negative 1$\sigma$ quantiles, respectively.
The complete movie of the Quantile map covering the full range of net count of photons in simulated spectra is presented as on line material.}
\label{fig8}
\end{figure*}

\begin{figure*}
\centering
\includegraphics[width=11cm,angle=0]{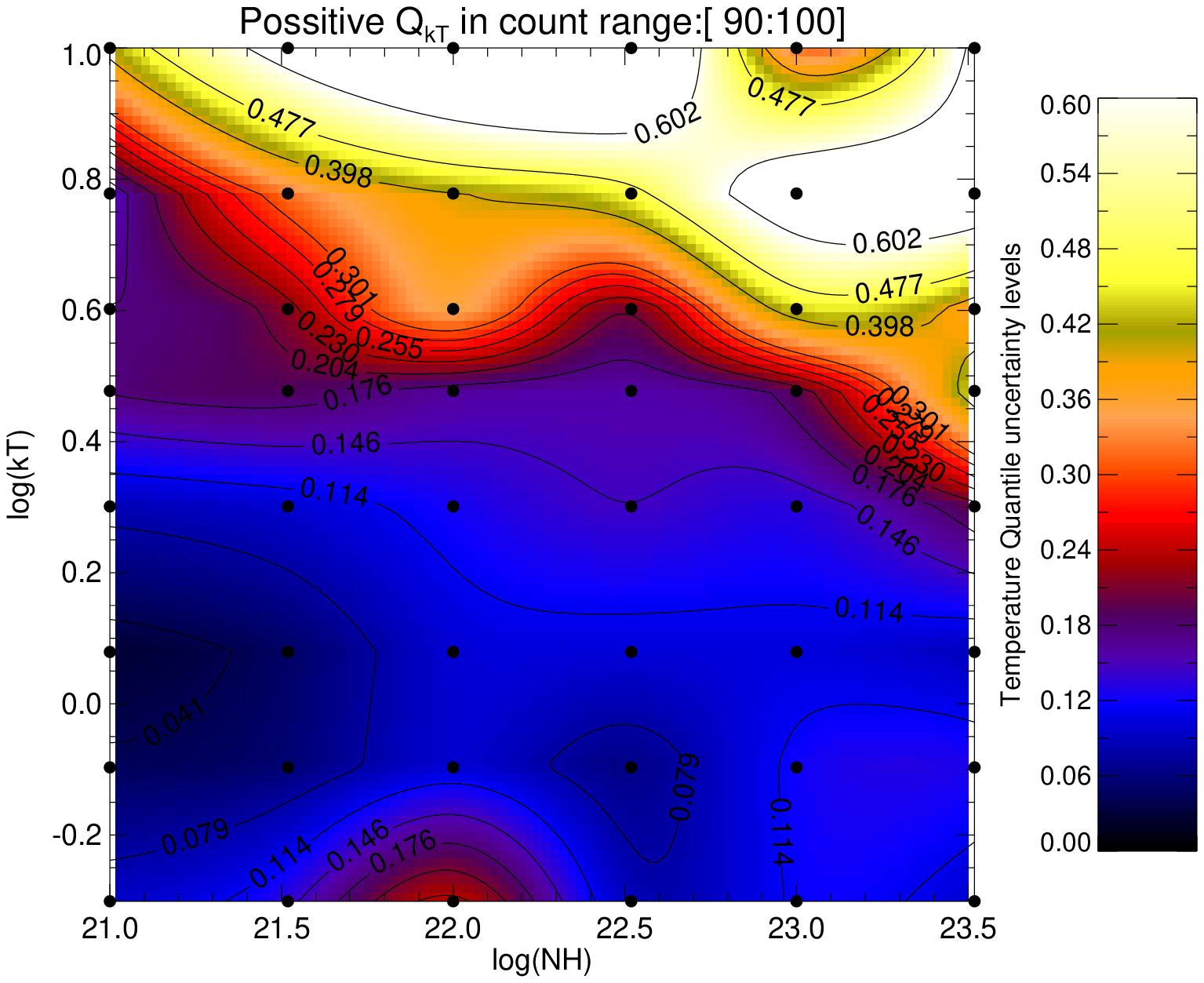}
\includegraphics[width=11cm,angle=0]{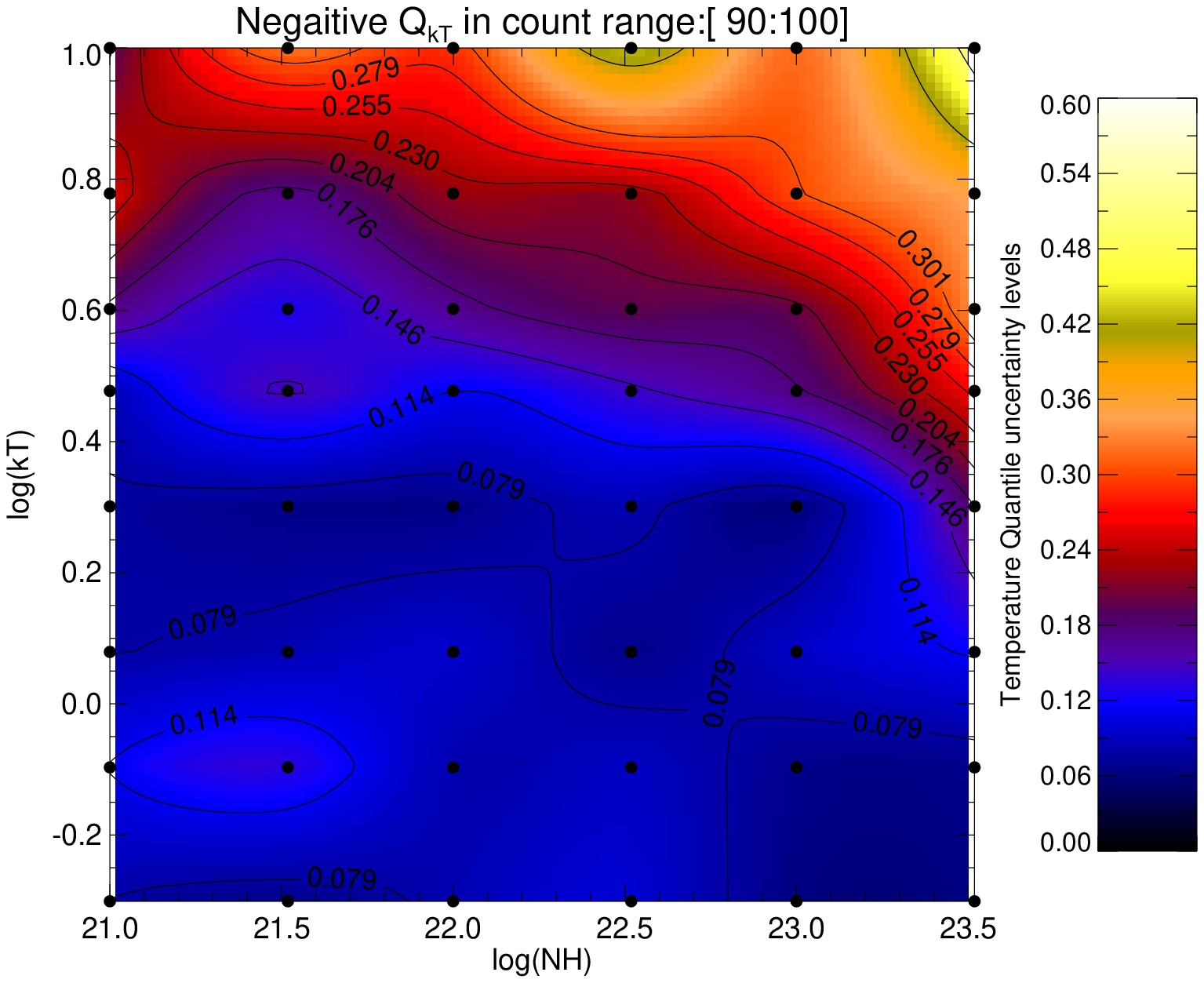}\\
\caption{\small \it 2D-temperature example Quantile maps from simulations of thermal models with 90 to 100 net photon counts.  Left and 
right columns refer to positive and negative 1$\sigma$ quantiles, respectively.
The complete movie of the Quantile map covering the full range of net count of photons in simulated spectra is presented as on line material.}
\label{fig9}
\end{figure*}

\begin{figure*}
\centering
\includegraphics[width=11cm,angle=0]{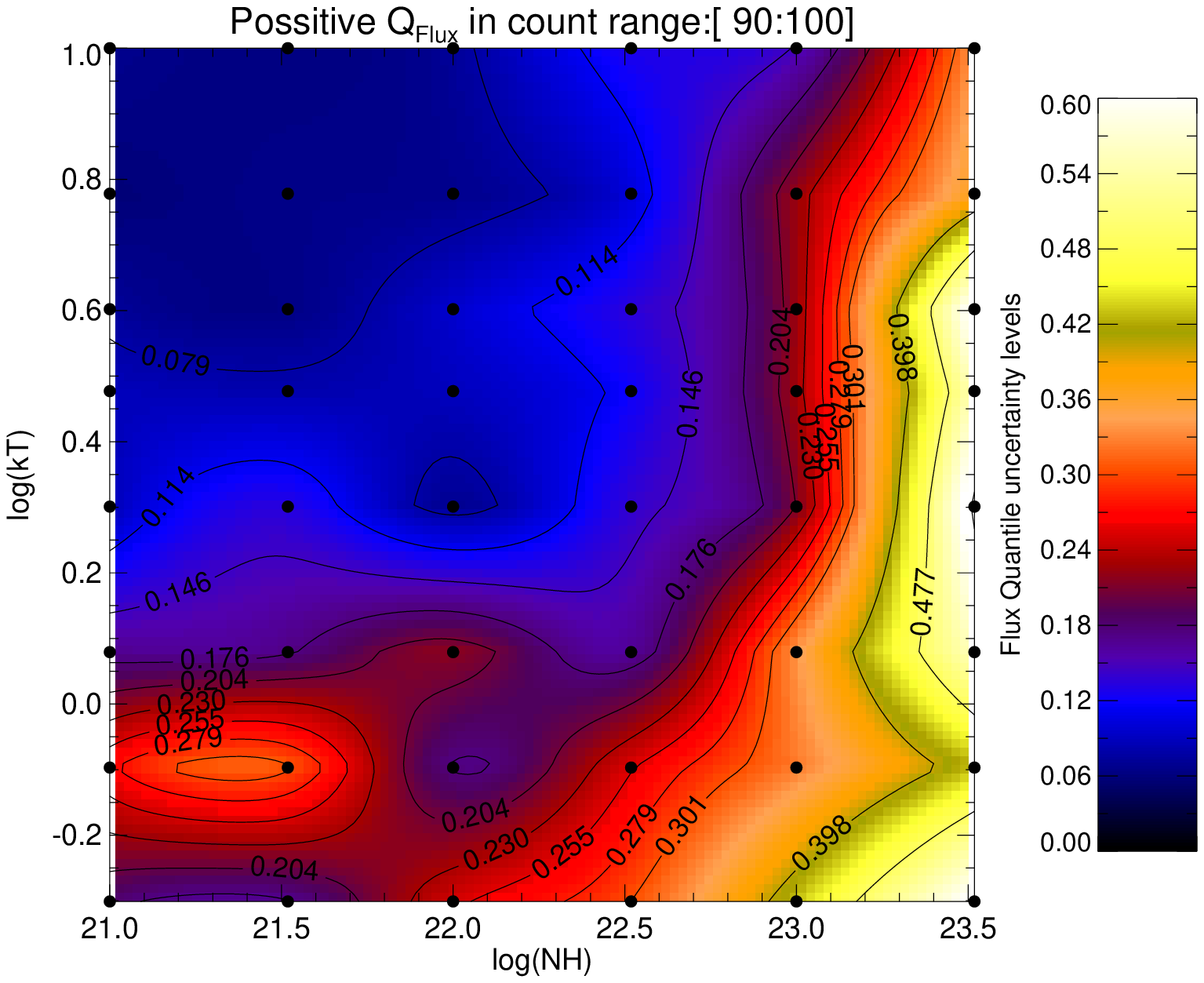}
\includegraphics[width=11cm,angle=0]{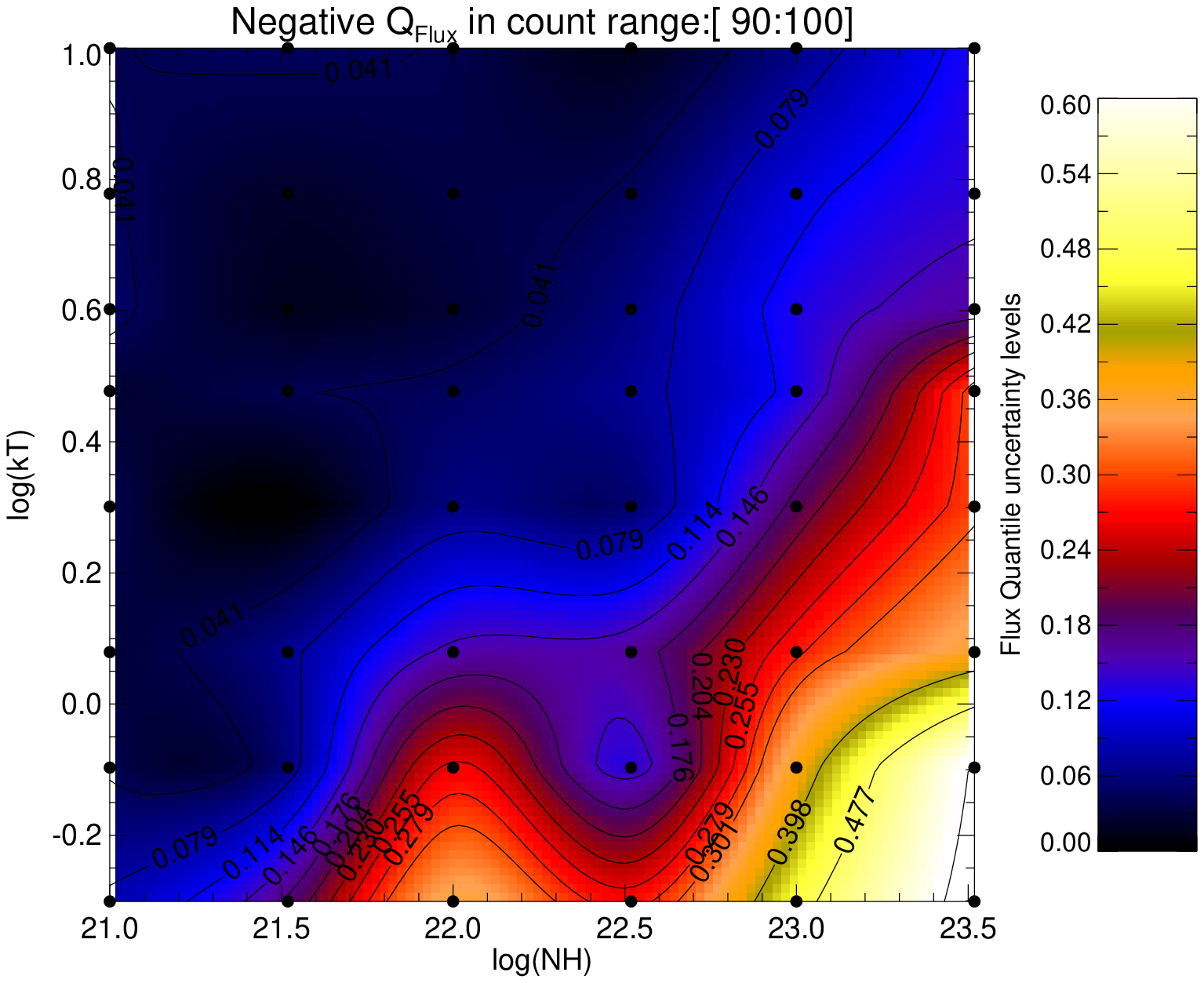}\\
\caption{\small \it 2D-Flux example Quantile maps from simulations of thermal models with 90 to 100 net photon counts.  Left and 
right columns refer to positive and negative 1$\sigma$ quantiles, respectively.
The complete movie of the Quantile map covering the full range of net count of photons in simulated spectra is presented as on line material.}
\label{fig10}
\end{figure*}

\begin{figure*}
\centering
\includegraphics[width=11cm,angle=0]{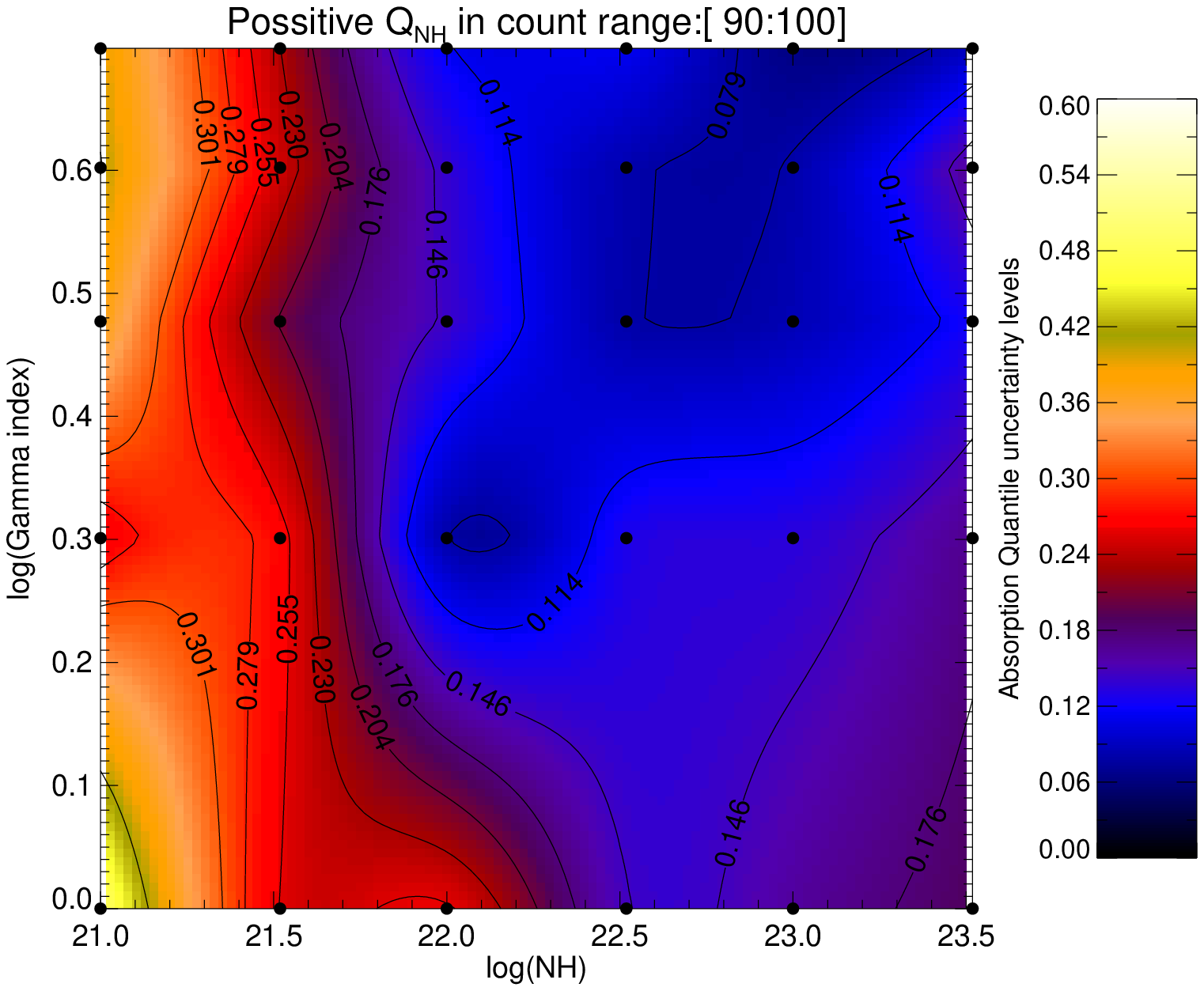}
\includegraphics[width=11cm,angle=0]{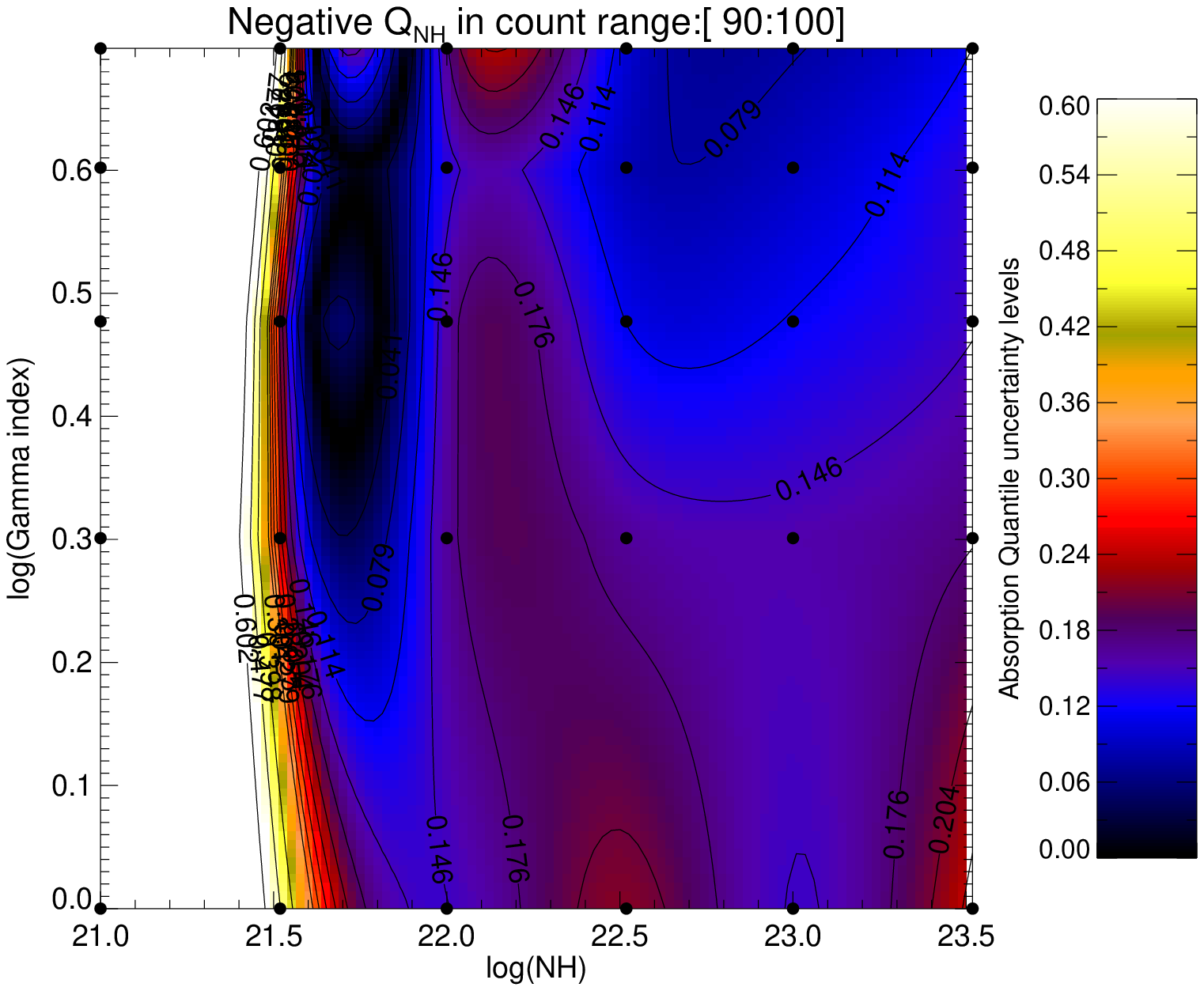}\\
\caption{\small \it 2D-N$_{\rm H}$ example Quantile maps from simulations of non-thermal models with 90 to 100 net photon counts. Left and 
right columns refer to positive and negative 1$\sigma$ quantiles, respectively.
The complete movie of the Quantile map covering the full range of net count of photons in simulated spectra is presented as on line material.}
\label{fig11}
\end{figure*}

\begin{figure*}
\centering
\includegraphics[width=11cm,angle=0]{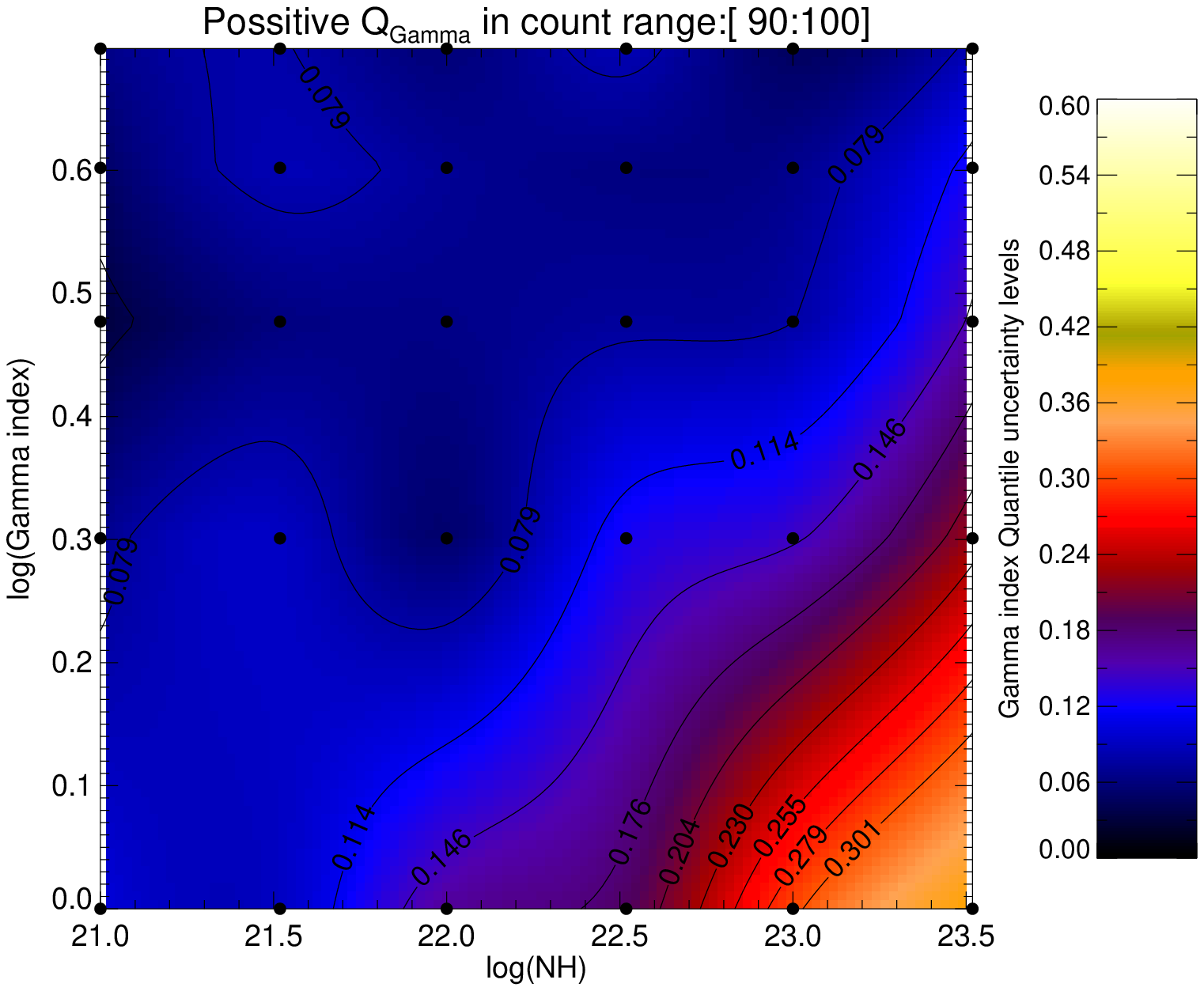}
\includegraphics[width=11cm,angle=0]{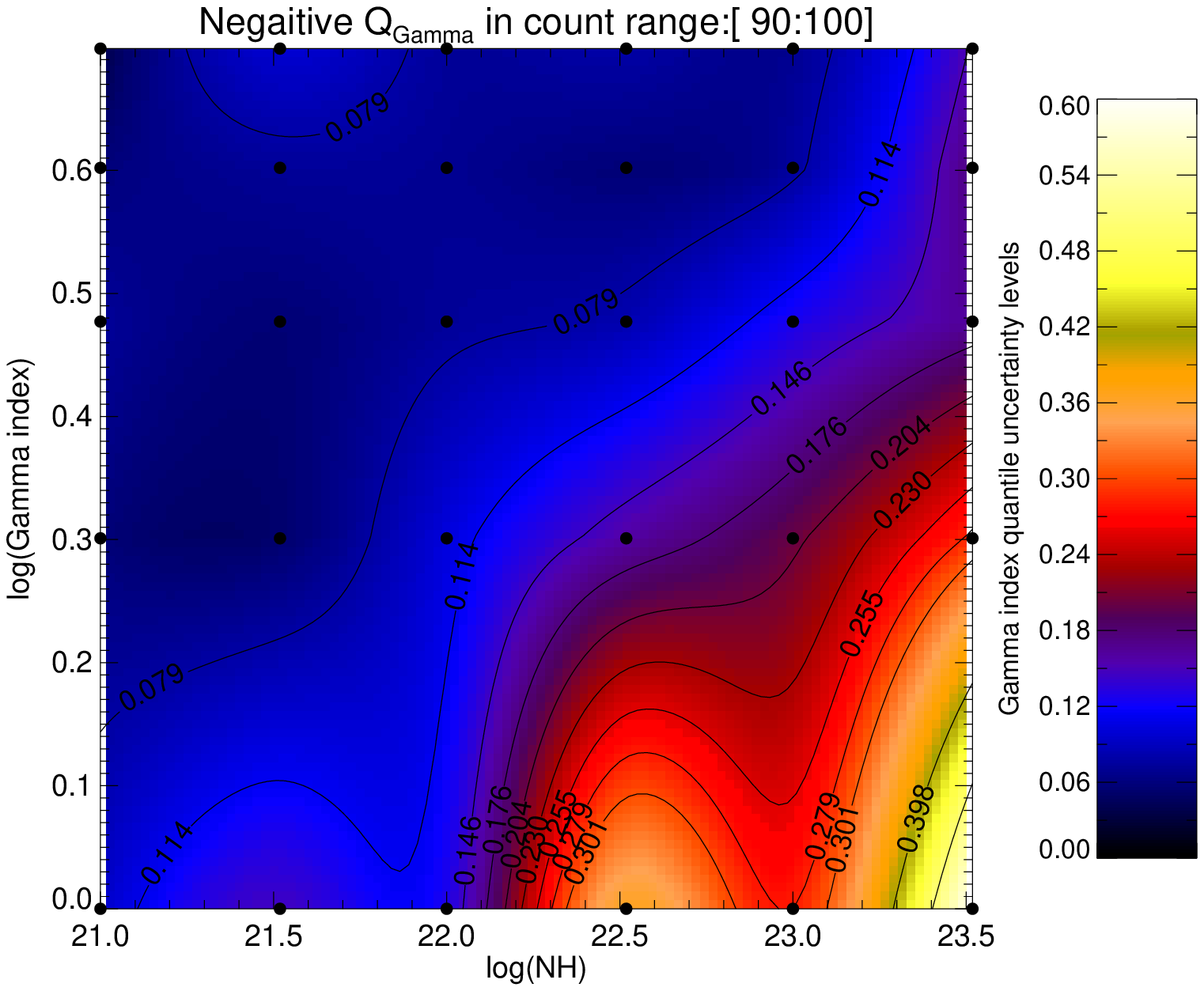}\\
\caption{\small \it 2D-$\Gamma$ index example Quantile maps from simulations of non-thermal models with 90 to 100 net photon counts. Left and 
right columns refer to positive and negative 1$\sigma$ quantiles, respectively.
The complete movie of the Quantile map covering the full range of net count of photons in simulated spectra is presented as on line material.}
\label{fig12}
\end{figure*}

\begin{figure*}
\centering
\includegraphics[width=11cm,angle=0]{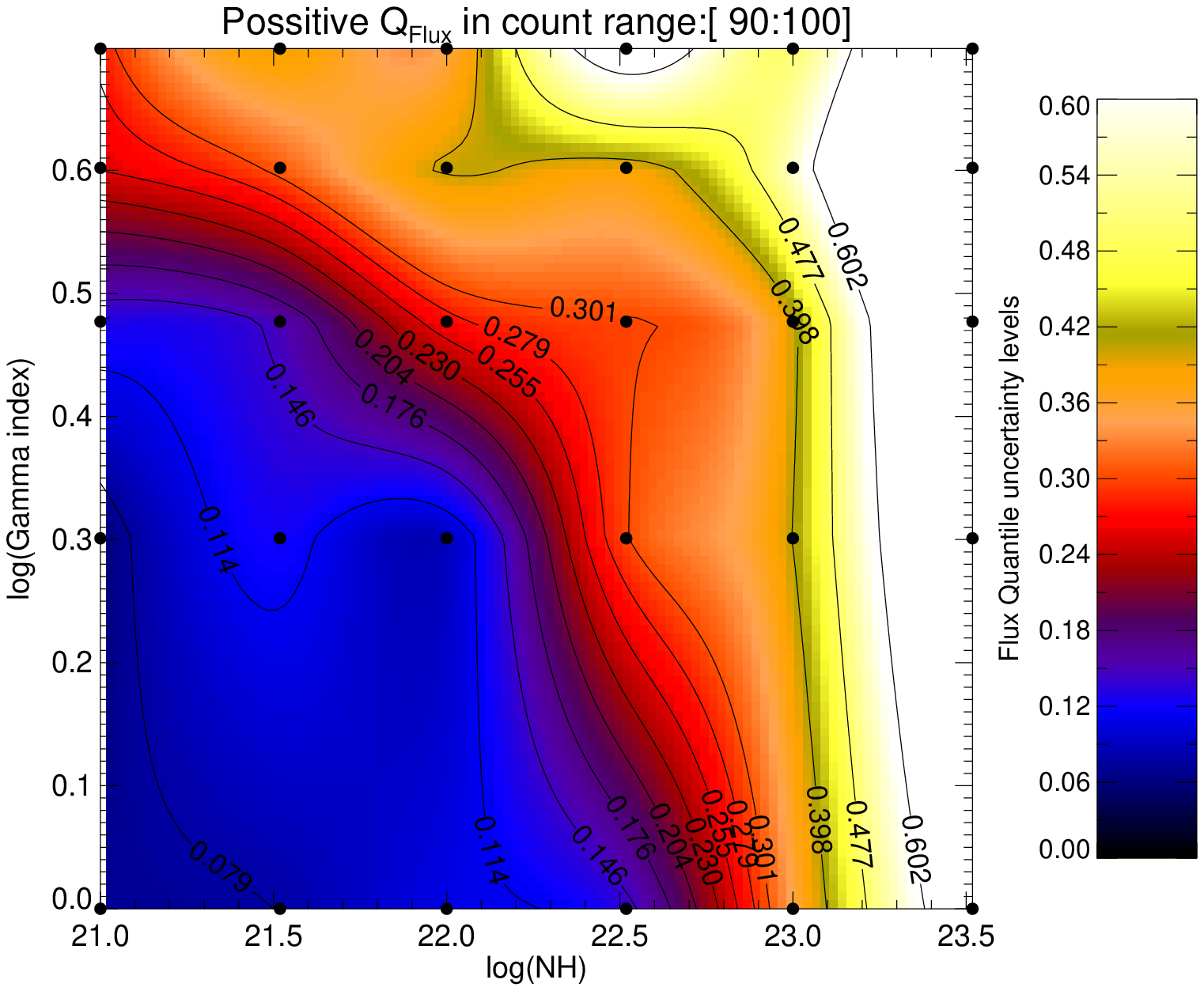}
\includegraphics[width=11cm,angle=0]{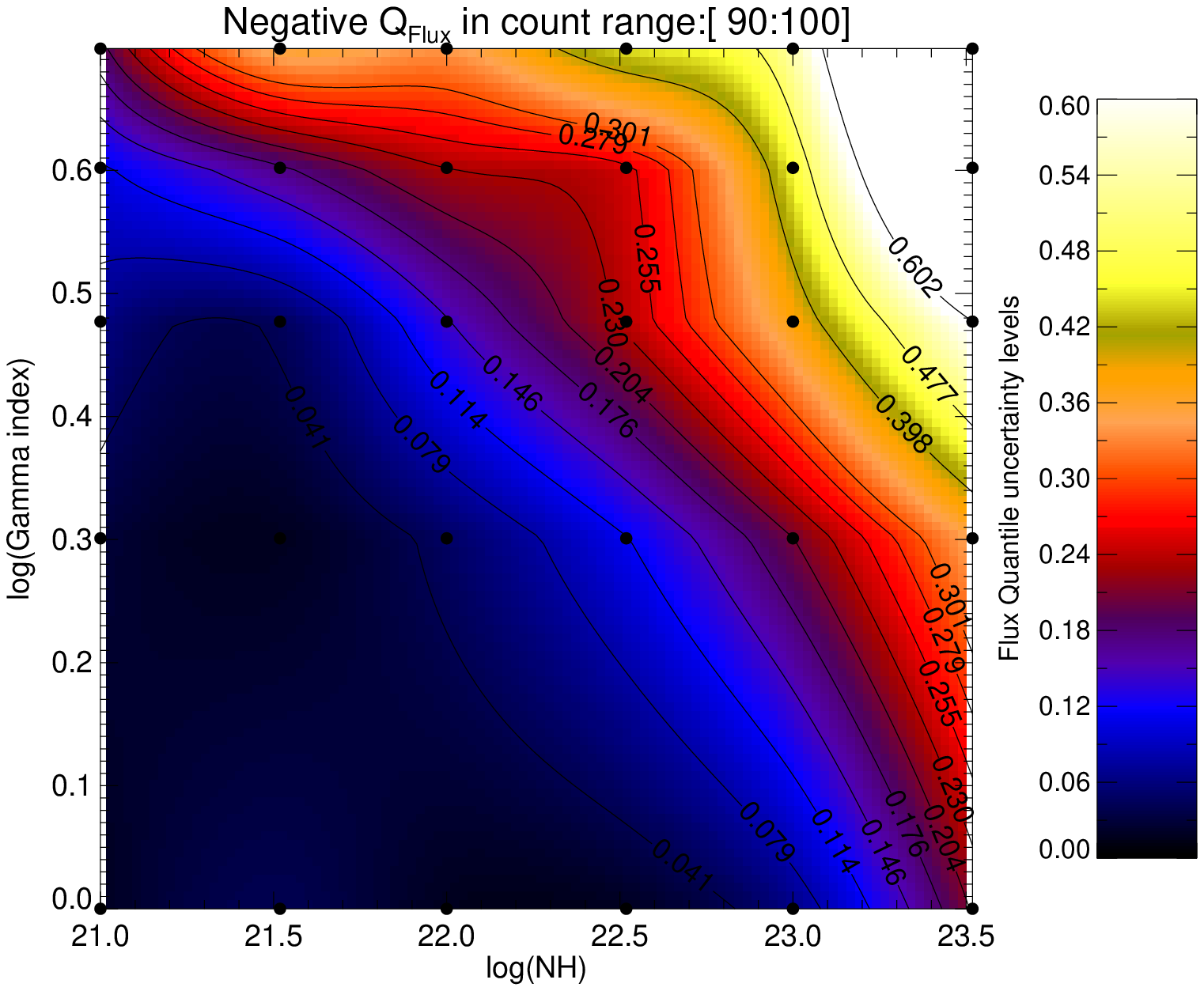}\\
\caption{\small \it 2D-Flux example Quantile maps from simulations of non-thermal models with 90 to 100 net photon counts. Left and 
right columns refer to positive and negative 1$\sigma$ quantiles, respectively.
The complete movie of the Quantile map covering the full range of net count of photons in simulated spectra is presented as on line material.}
\label{fig13}
\end{figure*}

\begin{table*}
\caption{Coefficients for correction of the absorption (N$_{\rm H}$) Q uncertainty in terms of the background contamination fraction for a thermal X-ray source model.}
\label{t:errorcoef}
\begin{center}
\begin{tabular}{lcccc}
\hline
 Model    & \multicolumn{1}{c}{20 ph} &\multicolumn{1}{c}{30 ph} & \multicolumn{1}{c}{100 ph} &\multicolumn{1}{c}{300 ph}  \\  
\cline{2-5}
1	&      0.56999999 ,      -0.79058786 &       0.29294601 ,      -0.35787199 &       0.18000001 ,      -0.32236677 &       0.11482079 ,     -0.087020800 \\ 
2	&       1.5800000 ,       -1.5800000 &       0.43000001 ,      -0.43000001 &       0.18000001 ,      -0.18000001 &      0.029938766 ,     -0.029938766 \\ 
3	&       1.0000000 ,       -1.0000000 &       0.59519789 ,      -0.59519789 &       0.11634470 ,      -0.11634470 &       0.10000000 ,      -0.10000000 \\ 
4	&     0.049715455 ,     -1.0000000 &       0.62929869 ,       -3.2919384 &       0.12753756 ,      -0.41111881 &    0.00016174661 ,      -0.17443736 \\ 
5	&      0.11171015 ,      -0.61089794 &      0.079572385 ,      -0.52813251 &      0.079557867 ,      -0.11617288 &      0.043629286 ,     -0.044922529 \\ 
6	&       1.3000000 ,       -2.0934750 &       0.67000002 ,      -0.75529767 &      0.093786779 ,      -0.25956268 &      0.086018041 ,     -0.057652803 \\ 
7	&      0.88999999 ,      -0.57335823 &       0.27781488 ,      -0.44000000 &       0.18754265 ,     -0.062686584 &       0.16403699 ,     -0.039999999 \\ 
8	&      0.58999997 ,      -0.60203861 &       0.41679498 ,      -0.35140125 &       0.42944980 ,      -0.12969924 &       0.28124141 ,     -0.022691981 \\ 
9	&      0.87000000 ,      -0.87000000 &       0.49000001 ,      -0.49000001 &       0.23999999 ,      -0.23999999 &       0.11832534 ,      -0.11832534 \\ 
10	&      0.38999999 ,     -0.790000 &       0.27743047 ,     -0.440000 &      0.063187085 ,      -0.28924141 &      0.034755657 ,     -0.029999999 \\ 
11	&      0.63999999 ,      -0.64599982 &       0.10191273 ,      -0.61887488 &       0.11436878 ,      -0.18611558 &      0.030987269 ,      -0.11433628 \\ 
12	&      0.38999999 ,      -0.69138395 &       0.28999999 ,      -0.37843546 &       0.21054822 ,      -0.20999999 &      0.067875243 ,      -0.19371328 \\ 
13	&      0.58999997 ,      -0.69999999 &       0.50999999 ,      -0.57290317 &       0.44196881 ,      -0.10809053 &       0.23690649 ,     -0.088281814 \\ 
14	&       1.0000000 ,       -1.0000000 &       0.23863081 ,      -0.23000000 &       0.16731354 ,     -0.096115480 &      0.029999999 ,     -0.087420713 \\ 
\hline
\end{tabular}
\end{center}
\label{slopes}
\end{table*}%

\begin{table*}
\caption{Coefficients for correction of the temperature (kT) Q uncertainty in terms of the background contamination fraction for a thermal X-ray source model.}
\label{t:errorcoef}
\begin{center}
\begin{tabular}{lcccc}
\hline
 Model    & \multicolumn{1}{c}{20 ph} &\multicolumn{1}{c}{30 ph} & \multicolumn{1}{c}{100 ph} &\multicolumn{1}{c}{300 ph}  \\  
\cline{2-5}
1	&       1.1791490 ,      -0.69999999 &        1.0236216 ,      -0.38275211 &       0.35172140 ,      -0.16000000 &      0.070000000 ,      -0.12034473 \\ 
2	&      0.59865859 ,      -0.40000001 &       0.15000001 ,      -0.22137435 &      0.079999998 ,      -0.10181730 &    -0.0058132137 ,     -0.020912962 \\ 
3	&       1.6737583 ,       -1.0000000 &       0.40438804 ,      -0.67145439 &       0.25332680 ,      -0.12570339 &      0.080397019 ,     -0.032017613 \\ 
4	&       1.5045421 ,      -0.19886096 &        1.0100729 ,      -0.70493358 &       0.35332554 ,      -0.19089511 &       0.12219090 ,     -0.064251043 \\ 
5	&       1.0285405 ,      -0.22003141 &        1.3540962 ,      -0.14531649 &       0.16706693 ,     -0.073932650 &      0.048124737 ,     -0.019780164 \\ 
6	&       1.7827508 ,       -1.7827508 &       0.77129511 ,      -0.77129511 &       0.13526567 ,      -0.13457852 &      0.020276173 ,      -0.14306573 \\ 
7	&       1.9000000 ,       -1.9000000 &        1.6500000 ,       -1.6500000 &        1.3150349 ,      -0.89999998 &       0.68300091 ,      -0.15976273 \\ 
8	&       1.6700000 ,      -0.56000000 &        1.5449302 ,      -0.46316759 &       0.43428638 ,      -0.35728049 &       0.12876973 ,      -0.17824362 \\ 
9	&       1.1222106 ,      -0.33227760 &        1.1170738 ,      -0.14316164 &       0.38273876 ,     -0.072976206 &      0.058509901 ,     -0.092050602 \\ 
10	&      0.98000002 ,      -0.44000000 &       0.68061929 ,      -0.27748997 &       0.54441994 ,      -0.16165912 &       0.21343224 ,      -0.15936723 \\ 
11	&      0.68000001 ,      -0.68000001 &       0.55418089 ,      -0.19408860 &       0.69571872 ,      -0.22827677 &       0.13584275 ,      -0.12354602 \\ 
12	&       1.7900000 ,      -0.28999999 &        1.3700000 ,      -0.25924568 &        1.2029648 ,      -0.14825462 &       0.36546867 ,     -0.050000001 \\ 
13	&      1.90000 ,      -0.70999998 &       1.70000 ,      -0.52999997 &        1.5044654 ,      -0.41568087 &       0.48996156 ,      -0.33525367 \\ 
14	&      1.14000 ,      -0.63999999 &      0.990000 ,      -0.34000000 &       0.80259332 ,      -0.19586915 &        1.2665589 ,      -0.15802399 \\ 
\hline
\end{tabular}
\end{center}
\label{slopes}
\end{table*}%

\begin{table*}
\caption{Coefficients for correction of the unabsorbed flux (flux) Q uncertainty in terms of the background contamination fraction for a thermal X-ray source model.}
\label{t:errorcoef}
\begin{center}
\begin{tabular}{lcccc}
\hline
 Model    & \multicolumn{1}{c}{20 ph} &\multicolumn{1}{c}{30 ph} & \multicolumn{1}{c}{100 ph} &\multicolumn{1}{c}{300 ph}  \\  
\cline{2-5}
1	&     0.600000 ,     -0.330000 &       0.54025202 ,      -0.11000000 &       0.10682827 ,     -0.050495573 &      0.040233470 ,     -0.036603905 	\\
2	&      0.60000002 ,      -0.22000000 &       0.26444894 ,     -0.075128038 &       0.14690279 ,     -0.035009770 &     0.0088736486 ,    -0.0088736486	\\
3	&      0.61000001 ,     -0.330000 &       0.46442869 ,     -0.022708227 &      0.091840132 ,     -0.074119511 &      0.033806779 ,     -0.041856700	\\
4	&      0.32547463 ,       0.22537710 &       0.71408237 ,    -0.0064318752 &      0.094341279 ,     -0.068698972 &    -0.0054402611 ,     -0.019880270	\\
5	&      0.63611881 ,      -0.58999997 &       0.38808347 ,      -0.36155323 &       0.21717026 ,      -0.21975059 &       0.16171609 ,     -0.074638882	\\
6	&      0.69999999 ,      -0.69999999 &       0.55000001 ,      -0.55000001 &       0.23215765 ,     -0.097066930 &       0.15446032 ,     -0.041110827	\\
7	&       1.8000000 ,       -1.8000000 &        1.3300000 ,       -1.3300000 &       0.62809720 ,      -0.20183422 &       0.38454859 ,      -0.19000000	\\
8	&       1.8800000 ,      -0.49000001 &        1.6358418 ,      -0.24090247 &       0.83024481 ,      -0.15000001 &       0.28627373 ,      -0.12124413	\\
9	&      0.33069971 ,      -0.33069971 &       0.22148313 ,      -0.22148313 &      0.077915734 ,     -0.062867283 &      0.024395005 ,     -0.040764767	\\
10	&      0.38207975 ,      -0.38999999 &       0.28651545 ,     -0.070000000 &      0.072748663 ,     -0.029204307 &      0.020000000 ,     -0.063466439	\\
11	&      0.27969115 ,      -0.27969115 &       0.20694499 ,     -0.060739656 &      0.086205055 ,     -0.049771037 &    0.00058630777 ,     -0.068272805	\\
12	&      0.38270376 ,      -0.38999999 &       0.29421943 ,      -0.27000001 &       0.12630442 ,      -0.19000000 &      0.029999999 ,      -0.13910898	\\
13	&     0.900000 ,      -0.41000000 &      0.770000 ,      -0.28565844 &       0.63059201 ,     -0.060723811 &       0.23391268 ,     -0.070676279	\\
14	&       1.3000000 ,       -1.3000000 &       0.60592199 ,      -0.80000001 &       0.51963257 ,      -0.23999999 &      0.039999999 ,      -0.15035874	\\
\hline
\end{tabular}
\end{center}
\label{slopes}
\end{table*}%

\begin{table*}
\caption{Coefficients for correction of the Absorption (NH) Q uncertainty in terms of the background contamination fraction for a non-thermal X-ray source model.}
\label{t:errorcoef}
\begin{center}
\begin{tabular}{lccc}
\hline
 Model    & \multicolumn{1}{c}{30 ph} & \multicolumn{1}{c}{100 ph} &\multicolumn{1}{c}{300 ph}  \\  
\cline{2-4}
1	&      0.25650181 ,       -1 &       0.12430397 ,      -0.50202637 &       0.18169266 ,       -1.1245344\\ 
2	&      0.40523864 ,      -0.30208423 &       0.11622960 ,       -1.1318363 &       0.28298807 ,       -2.9794969\\ 
3	&      0.22044614 ,       -1 &       0.55048480 ,     -0.051447758 &       0.52473601 ,       -3.2887644\\ 
4	&     0.500000 ,       -4.6345525 &       0.16597746 ,      -0.97468140 &      0.066336864 ,      -0.31664817\\ 
5	&      0.12215644 ,      -0.79181550 &       0.10350067 ,      -0.29293987 &      0.074220414 ,      -0.16425698\\ 
6	&     0.089567717 ,       -1.0198450 &       0.23257742 ,      -0.25237991 &      0.096013813 ,      -0.17154169\\ 
7	&     0.570000 ,     -0.250000 &       0.57413525 ,      -0.34905480 &       0.37936515 ,      -0.28201250\\ 
8	&      0.27773648 ,       -1.2329168 &       0.36234157 ,      -0.15427702 &       0.19328749 ,      -0.14761850\\ 
\hline
\end{tabular}
\end{center}
\label{slopes}
\end{table*}%

\begin{table*}
\caption{Coefficients for correction of the Gamma index ($\Gamma$) Q uncertainty in terms of the background contamination fraction for a non-thermal X-ray source model.}
\label{t:errorcoef}
\begin{center}
\begin{tabular}{lccc}
\hline
 Model    & \multicolumn{1}{c}{30 ph} & \multicolumn{1}{c}{100 ph} &\multicolumn{1}{c}{300 ph}  \\  
\cline{2-4}
1	&      0.19699650 ,     -0.081710183 &      0.052672355 ,     -0.093241764 &      0.039347859 ,     -0.050306389 \\ 
2	&      0.25117132 ,      -0.10266247 &      0.078957645 ,     -0.067741463 &      0.038052451 ,     -0.060455948 \\ 
3	&      0.21329448 ,     -0.016207271 &       0.22695826 ,     -0.010517673 &       0.16103488 ,     -0.046310364 \\ 
4	&     0.400000 ,     -0.500000 &      0.097128085 ,      -0.33579614 &     0.0047083283 ,      -0.23929186 \\ 
5	&     0.082105410 ,      -0.29366887 &      0.091130471 ,      -0.13137739 &      0.040882831 ,     -0.059770894 \\ 
6	&      0.12310893 ,      -0.28904541 &       0.22853133 ,      -0.14899948 &       0.10236678 ,      -0.12266170 \\ 
7	&      1.50000 ,     -0.600000 &      0.800000 ,     -0.500000 &       0.35792970 ,      -0.52635195 \\ 
8	&      0.43658025 ,      -0.49066309 &       0.16724924 ,      -0.19066881 &      0.067044326 ,     -0.043196329 \\ 
\hline
\end{tabular}
\end{center}
\label{slopes}
\end{table*}%

\begin{table*}
\caption{Coefficients for correction of the unabsorbed flux (flux) Q uncertainty in terms of the background contamination fraction for a non-thermal X-ray source model.}
\label{t:errorcoef}
\begin{center}
\begin{tabular}{lccc}
\hline
 Model    & \multicolumn{1}{c}{30 ph} & \multicolumn{1}{c}{100 ph} &\multicolumn{1}{c}{300 ph}  \\  
\cline{2-4}
1	&      0.17029916 ,     -0.027228378 &      0.053110142 ,     -0.043069600 &      0.027653891 ,     -0.025060421\\  
2	&       1.0406915 ,      0.012650287 &      0.077507279 ,     -0.039221376 &      0.073060202 ,     -0.035090637\\  
3	&      0.77695532 ,      0.013296297 &       0.90757450 ,     -0.062870060 &       0.53249562 ,      -0.12403132\\  
4	&      0.23904451 ,     -0.040587606 &      0.048288826 ,     -0.046017530 &     0.0500000 ,     -0.041524861\\  
5	&      0.58708610 ,     -0.093046929 &       0.31389033 ,      -0.23559708 &       0.12070400 ,      -0.12857378\\  
6	&     0.540000 ,      -0.55078238 &        1.3023024 ,      -0.51431365 &       0.41070973 ,      -0.56508111\\  
7	&       1.1208722 ,      -0.10905732 &        1.1694971 ,     -0.072869982 &       0.38194822 ,      -0.13259251\\  
8	&       2.2073709 ,      -0.56064182 &        1.1706366 ,      -0.44397279 &       0.40848299 ,      -0.30326180\\  
\hline
\end{tabular}
\end{center}
\label{slopes}
\end{table*}%

\end{document}